%% file: article.tex
\documentclass[useAMS,usenatbib,usegraphicx]{mn2e}
\usepackage{amsmath,amssymb,amsbsy,amsfonts}
\usepackage[T1]{fontenc}
\usepackage[british]{babel}
\usepackage{mathptmx}
\DeclareMathAlphabet{\mathcal}{OMS}{cmsy}{m}{n}
\DeclareSymbolFont{greekletters}{OML}{cmm} {m}{it}
\DeclareMathSymbol{\epsilon}{\mathalpha}{greekletters}{"0F}
\DeclareMathSymbol{\theta}{\mathalpha}{greekletters}{"12}
\DeclareMathSymbol{\rho}{\mathalpha}{greekletters}{"1A}
\DeclareMathSymbol{\phi}{\mathalpha}{greekletters}{"1E}
\DeclareMathSymbol{\psi}{\mathalpha}{greekletters}{"20}
\DeclareMathSymbol{\varepsilon}{\mathalpha}{greekletters}{"22}
\DeclareMathSymbol{\vartheta}{\mathalpha}{greekletters}{"23}
\DeclareMathSymbol{\varrho}{\mathalpha}{greekletters}{"25}
\usepackage[dvipsnames]{xcolor}
\usepackage{url}
\usepackage{microtype}
\usepackage{mathrsfs}
\usepackage{array}
\usepackage{graphicx}
\usepackage{units}
\usepackage{hyperref}
\usepackage{enumitem}
\usepackage{soul}
\usepackage{pdflscape}
\usepackage{balance}
\definecolor{custom-blue}{RGB}{41,22,206}
\hypersetup{
  colorlinks   = true, 
  urlcolor     = custom-blue, 
  linkcolor    = custom-blue, 
  citecolor    = custom-blue  
  }

\newcommand{\eq}[1]{Eq.\,\eqref{#1}}
\newcommand{\eqs}[1]{Eqs.\,\eqref{#1}}
\newcommand{\eqp}[1]{Eq.\,\ref{#1}}
\newcommand{\cc}{\mathrm{c}}

\newcommand{\ud}{\mathrm{d}}
\newcommand{\G}{\mathrm{G}}
\newcommand{\en}{\mathcal{E}}
\newcommand{\lz}{\ell_z}
\newcommand{\rg}{r_{\rm g}}

\newcommand{\rp}{r_{\rm p}}
\newcommand{\rt}{r_{\rm t}}
\newcommand{\rin}{r_0}
\newcommand{\rst}{R_\star}
\newcommand{\mst}{M_\star}
\newcommand{\mbh}{M}
\newcommand{\rsun}{R_\odot}
\newcommand{\msun}{M_\odot}
\newcommand{\abh}{a^*}

\renewcommand{\theequation}{\thesection.\arabic{equation}}

\title[Tidal disruptions by rotating black holes]
{Tidal disruptions by rotating black holes: relativistic hydrodynamics with Newtonian codes}
\author[E.~Tejeda, E.~Gafton, S.~Rosswog \& J.~C.~Miller]{Emilio Tejeda$^{1,2}$, 
Emanuel Gafton$^{2,3}$, Stephan Rosswog$^{2}$ and John C.~Miller$^{4}$\thanks{E-mail 
etejeda@astro.unam.mx (ET); emanuel.gafton@astro.su.se (EG); stephan.rosswog@astro.su.se (SR); john.miller@physics.ox.ac.uk (JM).} \\ 
$^{1}$Instituto de Astronom\'{i}a, Universidad Nacional Aut\'{o}noma de M\'{e}xico, AP 70-263, Distrito Federal, 04510, Mexico\\
$^{2}$Department of Astronomy and Oskar Klein Centre, Stockholm University, AlbaNova, SE-10691 Stockholm, Sweden\\
$^{3}$Nordic Optical Telescope, Rambla Jos\'{e} Ana Fern\'{a}ndez P\'{e}rez, 7, ES-38711 Bre\~{n}a Baja, Spain \\
$^{4}$Department of Physics (Astrophysics), University of Oxford, Keble Road, Oxford OX1 3RH, UK}

\pagerange{\pageref{firstpage}--\pageref{lastpage}}

\begin{document}

\maketitle

\label{firstpage}

\begin{abstract}
We propose an approximate approach for studying the relativistic 
regime of stellar tidal disruptions by rotating massive black holes. 
It combines an exact relativistic description of the hydrodynamical 
evolution of a test fluid in a fixed curved spacetime with a 
Newtonian treatment of the fluid's self-gravity. Explicit 
expressions for the equations of motion are derived for Kerr 
spacetime using two different coordinate systems. We implement the 
new methodology within an existing Newtonian Smoothed Particle 
Hydrodynamics code and show that including the additional physics 
involves very little extra computational cost. We carefully explore 
the validity of the novel approach by first testing its ability 
to recover geodesic motion, and then by comparing the outcome of 
tidal disruption simulations against previous relativistic studies. 
We further compare simulations in Boyer--Lindquist and 
Kerr--Schild coordinates and conclude that our approach allows 
accurate simulation even of tidal disruption events where the star 
penetrates deeply inside the tidal radius of a rotating black hole. 
Finally, we use the new method to study the effect of the black 
hole spin on the morphology and fallback rate of the debris streams 
resulting from tidal disruptions, finding that while the spin 
has little effect on the fallback rate, it does imprint heavily on 
the stream morphology, and can even be a determining factor in the 
survival or disruption of the star itself. Our methodology is discussed 
in detail as a reference for future astrophysical applications.
\end{abstract}


\section{Introduction} \label{sec:intro}

Gravity is the power source behind the most luminous phenomena in the Universe. For 
example, the accretion of gas onto stellar-mass black holes (BHs) is thought to power the majority of gamma-ray bursts (e.g.~\citealp{piran05a,meszaros06,lee07,nakar07}; \citealp*{gehrels09}), while gas being swallowed
by supermassive BHs provides the power supply for  active galactic nuclei \citep[e.g.][]{rees78,sanders88}. 
The vast majority of supermassive BHs however, go through extended ``dormant'' periods \citep{frank76,lidskii79} where they are starved of gas to accrete, but during which they can occasionally come back to life when a star passes by closely enough to be tidally disrupted \citep{rees88}. Such a tidal disruption event (TDE) delivers a substantial 
fraction of a stellar mass to the BH, and therefore provides a huge energy reservoir of 
$\sim 10^{53}$ erg ($M_{\rm acc}/0.1\,\msun$) that can potentially be tapped to 
power a flare of electromagnetic radiation.

\begin{figure*}
\includegraphics[width=\linewidth]{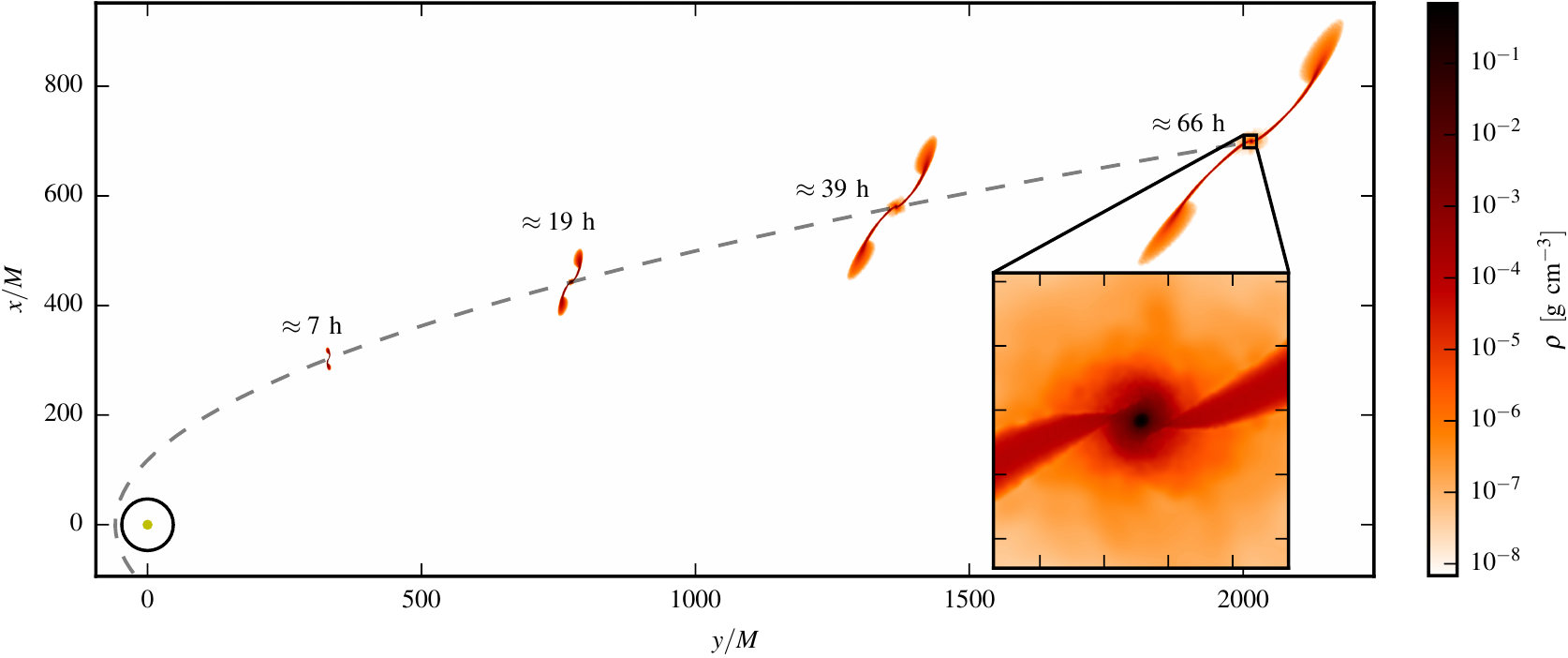}
\caption{Illustration of the geometric complexity of a tidal disruption event. The stellar matter density of a 1~$\msun$ star is shown at various times (as marked on the plot) after its partial disruption by a non-rotating  $10^6~\msun$ BH (impact parameter $\beta=0.8$, see text). The trajectory of the stellar centre of mass is shown as a dashed grey line. After the first passage, the star has been elongated into a very narrow curved cylinder with lobes at both ends, and a stellar core in the centre. At this stage, the debris is still receding from the BH; much later, the part ahead of the core will return back to the BH, while the trailing part is unbound. To illustrate the range of scales involved, we also show (lower left corner) the tidal radius (black circle) and the BH's gravitational radius (small yellow filled circle).}
\label{fig:disruption_geometry}
\end{figure*}%

Several TDEs have been observed, mostly in X-rays (\citealp*{bade96}; 
\citealp{komossa99,gezari03}), UV \citep{gezari06,gezari08,gezari09} and some in optical 
archival searches \citep{vanVelzen11} and transient searches \citep{cenko12a,gezari12,gezari16,holoien14,holoien16}. Recently,
collimated jets from two TDE candidates have been discovered by the Swift satellite 
\citep{bloom11,levan11,zauderer11,cenko12b}.
These observations have sparked a flurry of TDE studies (see, e.g.~\citealp{decolle12}; \citealp*{macleod12}; \citealp{guillochon13,manukian13}; \citealp*{guillochon14}; \citealp{GT15,bonnerot15}; \citealp{guillochon15}; \citealp*{hayasaki15}).

TDEs may also serve as probes for revealing the presence of intermediate-mass BHs. For 
example, white dwarf stars (WDs) can only be disrupted by Schwarzschild BHs with masses below $\sim 2 \times 10^5\;\msun$, whereas for larger BH masses the WD will fall through the horizon without being disrupted.  
For BHs below this limit, the tidal compression in a deep enough encounter can trigger
a significant release of thermonuclear energy \citep{luminet89}, resulting in a peculiar, type-Ia-like supernova (\citealp*{rosswog08a,rosswog09}; \citealp{macleod15}).  Furthermore, TDEs can also constrain BH parameters such as the spin rate. A rapidly rotating BH can disrupt stars that would otherwise be swallowed whole by a non-rotating BH, with a difference in the threshold BH mass as large as a factor of 10 (see Fig.~\ref{fig:tidal_radius}, below). Accurately capturing the relativistic effects due to a rotating BH becomes crucial for modelling events such as ASASSN-15lh \citep{dong16} as recently discussed by \citealp{leloudas16}. 

While some aspects of a TDE can be described analytically to a reasonable accuracy
\citep*{carter85,luminet86,rees88,stone13}, the disruption itself is 
a highly non-linear interaction between (relativistic) gravity (\citealp{kochanek94,stone12a,haas12,kesden12}; \citealp*{dai13,hayasaki13}; \citealp{cheng14,GT15,shiokawa15,guillochon15}), 
gas dynamics \citep*{evans89,ayal00,lodato09,guillochon13}, radiation (\mbox{\citealp{guillochon09}}; \mbox{\citealp{sadowski16a}}),
possibly magnetic fields \citep{sadowski16b} and, in extreme cases, tidally triggered thermonuclear reactions \citep{luminet89,rosswog08a,rosswog09,macleod15}. Therefore, the only way
to realistically study the disruption process itself is via numerical simulations. 

Independently of the complexity of the physical processes, the numerical simulation of a TDE poses a formidable challenge in itself due to the debris geometry and the length- and time-scales involved.
To fix ideas, consider the situation sketched in Fig.~\ref{fig:disruption_geometry}, where a 1~$\msun$ solar-type
star has just passed by a non-rotating $\mbh=10^6~\msun$ BH. In broad terms, the star is expected to be tidally disrupted if its periapsis distance $\rp$ lies within the tidal radius,
\begin{equation}
\begin{split}
\rt & \equiv \rst\left(\frac{\mbh}{\mst}\right)^{1/3}\\
& \simeq 48\,\rg\left(\frac{\rst}{\rsun}\right)
\left(\frac{\mst}{\msun}\right)^{-1/3}\left(\frac{\mbh}{10^6\;\msun}\right)^{-2/3},
\end{split}
\label{e0.1}
\end{equation}
where $\rg\equiv \G\mbh/\cc^2$ is the gravitational radius of the BH.\footnote{It is important to note here that $\rt$ as defined in \eq{e0.1} provides only a rough Newtonian estimate for the actual radius at which disruption takes place. This definition completely ignores relativistic effects (including BH rotation) which we will be discussing extensively in the rest of this paper.} The tidal radius is marked in the figure as a black, open circle; the BH's gravitational radius is marked with the small yellow filled circle. In this example, the star has passed the BH with an impact parameter of $\beta\equiv \rt/\rp= 0.8$.

In order to be able to use a spherically symmetric stellar equilibrium model as the initial condition, one must start such a simulation with an initial separation between the star and the BH greater than
\begin{equation}
\rin\sim 5\;\rt\simeq 120\;\rg\left(\frac{\rst}{\rsun}\right)
\left(\frac{\mst}{\msun}\right)^{-1/3} \left(\frac{\mbh}{10^6\;\msun}\right)^{-2/3},
 \end{equation}
and in this way assure that the tidal acceleration ($a_{\rm tid}\sim \G\mbh\rst/\rin^3$) is less than one per cent of the self-gravity and pressure forces inside the star (both of order $a_{\rm sg}\sim \G\mst/\rst^2$ as follows from consideration of hydrostatic equilibrium). It is easy to see that the  initial ratio of the accelerations $a_{\rm tid}/a_{\rm sg}$ at $\rin$ is approximately equal to $(\rin/\rt)^{-3}$, hence $\lesssim 1\%$ for a ratio of 5. Under these conditions, the initial stellar gas distribution only covers a small fraction of the computational volume,
\begin{equation}
\frac{{\rst}^3}{{\rin}^3}\simeq 10^{-8} \left(\frac{\mst}{\msun}\right)\left(\frac{10^6\;\msun}{\mbh} \right).
\end{equation}
While this does not pose a major challenge for a Lagrangian method like SPH \citep{monaghan05,rosswog09b,springel10a,rosswog15c}, it is a serious hurdle for Eulerian  methods, where vacuum is usually treated as a low-density background gas that must be evolved in the simulation.  Therefore, such simulations are often performed in the reference frame of the stellar centre of mass, with the BH  being treated as a time-varying, external force \citep{guillochon09,guillochon13}. This allows, on the one hand, reduction of the computational volume and, on the other, avoidance of excessive numerical advection error due to high velocity motion with respect to the computational grid.
 
 An additional challenge comes from the fact that the relevant signal velocity that enters the  
 CFL stability criterion (\citealp*{CFL28}) in a relativistic hydrodynamics simulation is the speed of light, so that the numerical time step is restricted to 
 \begin{equation}
 \Delta t <0.02  \; {\rm s} \left(\frac{\Delta x} {\rst/100}\right),
 \end{equation}
 where $\Delta x$ symbolises the smallest length-scale that needs to be resolved. This restriction may be relaxed after a disruption has occurred, but if the encounter is only weak and a stellar core survives, as in the example of Fig.~\ref{fig:disruption_geometry}, similar time step restrictions still apply after the encounter.  Therefore, a full simulation -- starting from several tidal radii and following the spreading of the stellar debris to large distances, the return of a fraction of the debris to the BH, and the subsequent circularization and formation of an accretion disc -- is prohibitively expensive for a fully relativistic treatment. Fixed metric approaches, on the other hand,  can obviously 
only be applied for phases where the self-gravity of the stellar debris can be safely ignored, and not for a full beginning-to-end simulation.
 
 For the numerical study of a TDE, this leaves the following options: 
 \begin{itemize}[noitemsep,topsep=0pt,leftmargin=0pt,itemindent=0pt,align=left]
 \item[a)] use an entirely Newtonian approach and restrict the focus to encounters that can be treated as non-relativistic with a reasonable accuracy  \citep{guillochon13,guillochon14,coughlin15};
\item[b)] use  a Newtonian hydrodynamics scheme together with a pseudo-Newtonian potential for approximately capturing some relativistic effects
 \citep{rosswog09,tejeda13a,hayasaki13,bonnerot15};
\item[c)] follow a post-Newtonian approach  for mildly relativistic encounters 
\citep{ayal00,ayal01,hayasaki15};
\item[d)] use a full numerical relativity approach by solving the Einstein equations, and restrict the attention mainly to regions near the BH (e.g.~\citealp{east14}); 
\item[e)] use a combination of some of the above approaches, as was recently done by, e.g., \cite{shiokawa15} and \cite{sadowski16b}.
\end{itemize} 

Clearly, each of the above approaches has its own shortcomings. Pseudo-Newtonian potentials, for instance, are usually tuned to reproduce special properties for the motion around a BH, but cannot reproduce all of the relevant relativistic effects simultaneously. Moreover, these kinds of potential have mostly been developed for non-rotating BHs \citep[see e.g.][for a comparison of some of the most commonly used pseudo-Newtonian potentials]{tejeda13a}, and they have been less successful in modelling (the more realistic) rotating BHs.

Post-Newtonian approaches, on the other hand, are computationally very demanding since they require the solution
of several Poisson equations (nine for the full approach of \citealp{ayal01}), while being unnecessary far from the BH and inaccurate close to it.  In addition, the computational burden for solving the Poisson equations seriously restricts the numerical resolution that can be afforded for the hydrodynamics.

In this paper, we suggest a hybrid approach that combines an exact relativistic treatment of the acceleration from a rotating BH with a Newtonian treatment of the fluid's self-gravity. We work out explicitly the accelerations in both Boyer--Lindquist (BL) and Kerr--Schild (KS) coordinates. The resulting equations are simple to implement within  Newtonian hydrodynamic codes, as we demonstrate here using the Newtonian SPH code described in detail in \cite{rosswog08b}. Since the fluid's contribution to the spacetime geometry is neglected, this approach is, of course, not entirely self-consistent. Nevertheless, as we will show below, it is exact far from the BH and reproduces known results to a high accuracy even for very deep encounters, while only minimally increasing the computational burden with respect to a Newtonian simulation. 

Although this new tool has mostly been developed for the study of TDEs, one can straightforwardly apply it in situations where self-gravity only impacts on the fluid, while the spacetime geometry is, to a good approximation, only determined by the BH.

The paper is organised as follows. In Sec.~\ref{sec:rel_effects} we discuss the most salient relativistic effects that are relevant for the study of TDEs. In Sec.~\ref{sec:coord_vol} we present the methodology used in this work for treating the exact relativistic evolution of a fluid in a curved spacetime coupled with an approximate treatment of the fluid's self-gravity. In Sec.~\ref{sec:validation} we present a number of tests designed to compare our method with known analytic solutions as well as with results of previous relativistic simulations. In Sec.~\ref{sec:app} we demonstrate the use of our methodology for studying TDEs involving a rotating BH. A summary of our method and results is given in  Sec.~\ref{sec:summary}. In the accompanying Appendix, we have collected explicit expressions for the fluid accelerations in Kerr spacetime together with a brief summary of particle motion in this spacetime.

\setcounter{equation}{0}
\section{Relativistic effects relevant for TDEs}
\label{sec:rel_effects}

In this section we summarise a number of length scales and relativistic effects that are relevant for TDEs, and discuss the regimes in which the effects are significant.

\vspace{-12pt}
\subsubsection*{Event horizon}
This can be thought of as a one-way membrane that matter and light can only cross going inwards. Since matter plunging into the event horizon becomes causally disconnected from the rest of the universe, the existence of an event horizon directly affects the overall dynamics and energy budget in an accretion system. For a Kerr BH with specific angular momentum $a$, the event horizon is located at\footnote{Here and for the rest of this work we adopt geometric units, with $\G=\cc=1$.}
\begin{equation}
r_\mathrm{eh} = \mbh + \sqrt{\mbh^2 - a^2}.
\label{reh}
\end{equation}
In what follows, the spin parameter $a$ is taken to be positive when referring to a BH co-rotating with the orbiting matter and negative when it is counter-rotating. 

\vspace{-12pt}
\subsubsection*{Innermost stable circular orbit}
This marks the transition radius within which stable circular motion is no longer possible. For a standard thin accretion disc, this implies the existence of an inner edge from which the fluid falls essentially freely into the BH. The radius of this orbit is a function of the spin parameter of the BH. We will not write here the well-known formula for it \citep[see e.g.][]{novikov} since it is rather complicated and we will not be using it in the following.

\vspace{-12pt}
\subsubsection*{Marginally bound circular orbit}
In general relativity there is a critical value for the angular momentum of a test particle below which the resulting centrifugal repulsion is not enough to prevent the trajectory from plunging into the BH's event horizon. This translates into a minimum periapsis distance that a given trajectory can attain. In the case of marginally bound particles (i.e.~particles with parabolic-like energies), the corresponding radius is given by \citep*{bardeen72}
\begin{equation}
r_\mathrm{mb} = 2\mbh - a + 2\sqrt{\mbh(\mbh - a)}.
\label{rmb}
\end{equation}
In the context of TDEs, the different ways in which this radius and the tidal radius scale with the BH's mass ($r_\mathrm{mb}\propto M$ and $\rt\propto M^{1/3}$, respectively) imply that, for a given type of star, there exists a maximum possible value of $\mbh$ above which the star will be swallowed whole inside the BH horizon before being tidally disrupted, see Fig.~\ref{fig:tidal_radius} below. 

\vspace{-12pt}
\subsubsection*{Periapsis precession}
The relativistic trajectory of a bound particle around a BH does not close on itself as in the Newtonian case. To first order, the particle's trajectory can be described as an elliptical orbit whose periapsis is subject to a shift $\propto M/\rp$ per revolution. However, this shift becomes arbitrarily large as $\rp \rightarrow r_\mathrm{mb}$. Compared to a Newtonian encounter, this relativistic effect has a profound impact on deep TDEs,  causing the tidal stream to self-intersect at smaller radii, steeper angles and with larger relative velocities. This leads to the formation of a strong shock that efficiently dissipates kinetic energy, injects additional turbulence into the returning stream, and can potentially speed up the circularization of the debris and the formation of an accretion disc \citep[see e.g.][]{shiokawa15,bonnerot15,sadowski16b}.

\vspace{-12pt}
\subsubsection*{Orbital plane precession}
Particles orbiting around a rotating BH are no longer constrained to move within a single plane. As a first approximation for bound orbits, the motion of a test particle can be described in terms of a precessing orbital plane that oscillates between two extremes with a frequency approximately equal to the Lense--Thirring precession rate $\propto a M/\rp^3$. If, following a TDE, the cross section of the debris stream remains sufficiently thin, this additional precession may prevent stream self-intersection for several orbits, which can significantly delay the circularization of the debris \citep{guillochon15}. Independently of the precise details of the circularization process, it is expected that a TDE involving a rotating BH will in general produce a geometrically thicker accretion disc than one involving a non-rotating BH.

\vspace{-12pt}
\subsubsection*{Binding energy of circular trajectories} 
Particles on circular trajectories around a BH are more tightly bound to the central object than in the Newtonian case. For fluid moving round in an accretion disc, this extra budget of potential energy directly affects the total luminosity emitted from the accretion disc.

\vspace{-12pt}
\subsubsection*{Enhanced tidal field}
A deeper gravitational potential well implies a steeper potential gradient and, therefore, an enhanced tidal field. This is reflected in, for instance, stretching factors up to 25 times larger in the relativistic description of a $10^6\,\msun$ non-rotating BH as compared to the Newtonian case \citep[see, e.g.~Figure A1 of][]{GT15}.

\vspace{-12pt}
\subsubsection*{Parameter space for TDEs}
\begin{figure}
\includegraphics[width=\linewidth]{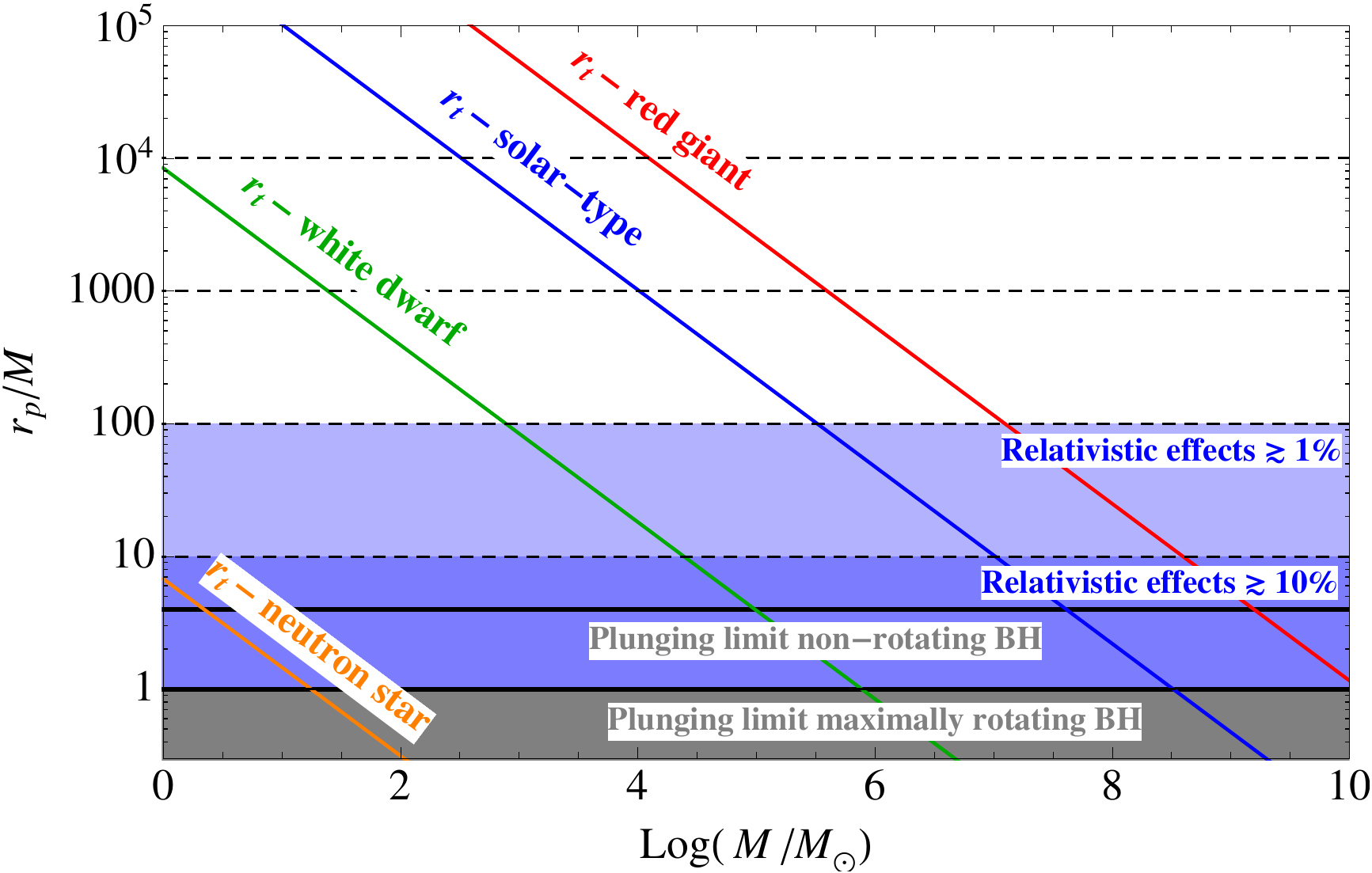}
\caption{Parameter space for the tidal disruption of different types of star. The vertical axis shows the periapsis distance of the parabolic trajectory along which a star approaches the central BH, whose mass $\mbh$ is shown on the horizontal axis. The solid, coloured lines indicate the tidal radii $\rt$ for different types of star. For a given type of star, we can expect to have a TDE only if the periapsis distance lies below the corresponding $\rt$ curve. The black horizontal lines represent the minimum possible periapsis distance of a parabolic trajectory, which corresponds to the radius of the marginally bound circular orbit and is equal to 4~$\mbh$ for a non-rotating BH and to 1~$\mbh$ for a maximally co-rotating BH. We also indicate the importance of relativistic effects by the blue-shaded regions (see main text). From top to bottom, the parameters chosen to represent each stellar type are: ($\mst=4\,\msun$, $\rst=18\,\rsun$), ($\mst=\msun$, $\rst=\rsun$), ($\mst=0.6\,\msun$, $\rst=0.015\,\rsun$), ($\mst=1.4\,\msun$, $\rst=10\,$km). }
\label{fig:tidal_radius}
\end{figure}

In Fig.~\ref{fig:tidal_radius} we have plotted the parameter space for TDEs in terms of the BH mass $\mbh$ and the periapsis distance $\rp$ for different types of star. For fixed BH mass and spin parameter, the lower limit for $\rp$ is given by $r_\mathrm{mb}$ as defined in \eq{rmb}. From this equation we get a minimum distance of 4~$\mbh$ for a non-rotating BH ($a=0$) and of 1~$\mbh$ for a maximally rotating BH ($a=\mbh$) when the star is in co-rotation with the BH. If, instead, the BH is counter-rotating, this distance becomes $\simeq 5.8\,\mbh$. The first two limits are indicated in Fig.~\ref{fig:tidal_radius} with thick horizontal lines. As $\rp \rightarrow r_\mathrm{mb}$, relativistic effects become increasingly important. In broad terms, relativistic corrections to quantities such as e.g.~binding energy or periapsis shift are of the order of a few percent for $10~\mbh<\rp<100~\mbh$, and can exceed $10\%$ for $\rp\lesssim 10~\mbh$. From this same figure we can see that, for instance, a main sequence star like the Sun can be disrupted by a central BH of at most $\mbh\sim 10^{7.5}\,\msun$ for a non-rotating BH and of $\mbh~\sim 10^{8.5}\,\msun$ for a maximally rotating one. It can further be seen that a TDE of a solar-type star takes place in the relativistic regime for $\mbh\gtrsim 10^{5.5}\,\msun$.

\setcounter{equation}{0}
\section{Coordinate time evolution of a perfect fluid}
\label{sec:coord_vol}

\subsection{Relativistic hydrodynamics equations}

Let us consider a perfect fluid evolving on a given curved spacetime described by the metric tensor $g_{\mu\nu}$. 
The state of a fluid element is characterised in terms of the rest mass density $\varrho$, the pressure $P$, the specific internal energy $u$, and the specific entropy $s$. Note that the rest mass density $\varrho$ is related to the baryon number density $n$ by $\varrho=m_0\,n$, where $m_0$ is the average baryonic rest mass of the fluid. All of these thermodynamical quantities are measured in the reference frame co-moving with the fluid element. Within the perfect fluid approximation, the specific entropy is a conserved quantity along the worldline of any given fluid element. The rest of the thermodynamic quantities are related through an equation of state. The motion of the fluid element itself is described by the four-velocity vector $U^\mu$ defined as the vector tangent to its worldline and normalised as\footnote{Here and in what follows we adopt Einstein's convention of summation over repeated indices, with Greek indices denoting spacetime components and Latin indices denoting only spatial components.} 
\begin{equation}
U^\mu\,U_\mu = g_{\mu\nu}U^\mu\,U^\nu = -1.
\label{e1.1}
\end{equation} 
The contravariant components of the four-velocity are given as $U^\mu = \ud x^\mu /\ud \tau$, where $\tau$ is the proper time of the fluid element and $(x^\mu) = (t,x^i)$ is a particular 3+1 splitting of the coordinate system where $t$ is the coordinate time and $x^i$ represents the spatial coordinates of the fluid element. 

The stress-energy tensor of the perfect fluid is expressed in terms of the above defined quantities as
\begin{equation}
T^{\mu\nu} = \varrho\,\omega\,U^\mu U^\nu + P\, g^{\mu\nu},
\label{e1.2}
\end{equation}
where $\omega = 1 + u + P/\varrho$ is the relativistic specific enthalpy. The fluid evolution equations follow from the conservation of baryon number 
\begin{equation}
\left(n\,U^\mu\right)_{;\mu} = 0,
\label{e1.3}
\end{equation}
and the local conservation of energy-momentum 
\begin{equation}
\left(T^{\mu\nu}\right)_{;\mu} = 0,
\label{e1.4}
\end{equation}
where a semicolon stands for covariant differentiation. These five equations (one in Eq.~\ref{e1.3} and four additional ones from the four components of Eq.~\ref{e1.4}), together with the fluid's equation of state, suffice for calculating the time evolution of a fluid described with the six variables  $(U^\mu,\varrho,u)$. 

By substituting \eq{e1.2} into \eq{e1.4}, employing the continuity equation \eqref{e1.3}, together with the first law of thermodynamics for a perfect fluid expressed in terms of the specific enthalpy as $\ud \omega = \ud P / \varrho$, we obtain the relativistic Euler equation
\begin{equation}
\frac{\ud U^\nu}{\ud \tau} = -\frac{1}{\varrho\,\omega}\frac{\partial P}{\partial x^\mu}\left(U^\mu U^\nu+g^{\mu\nu}\right) - \Gamma^\nu_{\lambda\mu}U^\lambda U^\mu,
\label{e1.4b}
\end{equation}
where $\Gamma^\nu_{\lambda\mu}$ denote the Christoffel symbols.

In the present work we are wanting to describe the evolution of the fluid in terms of the global time coordinate $t$ rather than the proper time $\tau$ which runs at different rates for fluid elements at different locations. For doing this, it is convenient to recast all of the proper time derivatives in \eqs{e1.3} and \eqref{e1.4b} as derivatives with respect to the coordinate time $t$ using the identity
\begin{equation}
\frac{\ud }{\ud t} = \frac{1}{\Gamma}\frac{\ud }{\ud \tau} = \frac{1}{\Gamma}\, U^\mu \frac{\partial }{\partial x^\mu},
\label{e1.5}
\end{equation}
where
\begin{equation}
\Gamma \equiv \frac{\ud t}{\ud \tau} = U^0
\label{e1.6}
\end{equation}
is a generalized Lorentz factor.\footnote{Note that this is a different quantity from the somewhat similar one (also denoted by $\Gamma$) used by \cite{HM66} and \cite{MW66} in considerations of spherical collapse, and subsequently used by other authors following on from them.} \eq{e1.1} can be used to rewrite \eq{e1.6} as
\begin{equation}
\Gamma = \left(-g_{\mu\nu}\, \dot{x}^\mu\, \dot{x}^\nu\right)^{-1/2},
\label{e1.7}
\end{equation}
where an overdot stands for the derivative with respect to the coordinate time $t$ and, clearly, $(\dot{x}^\mu) = (1,\ud x^i/\ud t)$.

With the aid of \eqs{e1.1}, \eqref{e1.5} and \eqref{e1.7}, we can use the time component of \eq{e1.4b} to rewrite its remaining spatial components as
\begin{equation}
\begin{split}
\frac{\ud^2 x^i}{\ud t^2} = & -\left(g^{i \lambda}-  \dot{x}^i g^{0 \lambda}\right)
\bigg[ \frac{1}{\Gamma^2\varrho\,\omega}\frac{\partial P}{\partial x^\lambda}  + \\
& \hspace{40pt}\left( \frac{\partial g_{\mu\lambda}}{\partial x^\sigma} - \frac{1}{2}\frac{\partial g_{\mu\sigma}}{\partial x^\lambda} \right)\dot{x}^\mu\, \dot{x}^\sigma \bigg].
\end{split}
\label{e1.8}
\end{equation}
On the other hand, if we compute the inner product of \eq{e1.4b} with the four-velocity and use the expressions in \eqs{e1.1}, \eqref{e1.5} and \eqref{e1.7} we get 
\begin{equation}
\frac{\ud u}{\ud t} = \frac{P}{\varrho^2} \frac{\ud \varrho}{\ud t},
\label{e1.9}
\end{equation}
which recovers again the first law of thermodynamics for a perfect fluid.
Finally, we can bring the expression for baryon number conservation, \eq{e1.3}, into the form
of an evolution equation by applying \eq{e1.5}:
\begin{equation}
\frac{\ud n}{\ud t} = -\frac{n}{\sqrt{-g}\,\Gamma}\frac{\partial}{\partial x^\mu}\left(\sqrt{-g}\,\Gamma\,\dot{x}^\mu\right),
\label{e1.10}
\end{equation}
where $g$ is the determinant of the four-metric $g_{\mu\nu}$.

Equations \eqref{e1.8} and \eqref{e1.9} are directly useful for our purposes since they can be straightforwardly implemented within our Newtonian SPH code. Equation \eqref{e1.10} however, is somewhat problematic because of the explicit time derivative of the Lorentz factor $\Gamma$ on the right hand side of the equation. Such time derivatives are known to limit the stability of relativistic hydrodynamics schemes, see for example  \cite{norman86}. We can avoid this difficulty altogether by introducing the auxiliary density variable 
\begin{equation}
 N = \sqrt{\frac{-g}{\gamma}}\,\Gamma\,n,
\label{e1.11} 
 \end{equation} 
where $\gamma$ is the determinant of the three-metric $\gamma_{ij} = g_{ij}$.  Using this new variable, \eq{e1.10} can be recast as 
\begin{equation}
\frac{\ud N}{\ud t} = - N\,\nabla_i \,\dot{x}^i - \frac{N}{\sqrt{\gamma}}\frac{\partial\sqrt{\gamma}}{\partial t},
\label{e1.12}
\end{equation} 
where $\nabla_i \,\dot{x}^i = 1/\sqrt{\gamma}\,\partial \left(\sqrt{\gamma}\,\dot{x}^i\right) /\partial x^i$ is the three-divergence as calculated on a spatial hypersurface $t=\textrm{const}$. Equation \eqref{e1.12} does not involve time derivatives of the Lorentz factor $\Gamma$. Furthermore, for the stationary metrics that we are interested in here -- such as the Kerr one -- the second term on the right hand side of \eq{e1.12} vanishes identically. 

In the following we shall use the SPH method for solving the hydrodynamic equations. Similarly to mass conservation in the 
Newtonian case, the SPH method allows one to either integrate a continuity equation (for mass or baryon number) or to enforce baryon conservation 
by simply keeping the mass/baryon number carried by each SPH particle fixed in time while the fluid is evolving. In this way, 
and in the special case of stationary metrics, we can altogether bypass the need to solve \eq{e1.12}. 
In order to see this, note that for stationary metrics we can rewrite \eq{e1.12} as
\begin{equation}
\frac{\partial N}{\partial t} = - \nabla_i (N\,\dot{x}^i),
\label{e1.13}
\end{equation} 
and then integrate \eq{e1.13} over any given spatial volume element $\mathcal{V}$. Doing so we obtain
\begin{equation}
\begin{split}
\frac{\ud}{\ud t} \mathcal{N} = \frac{\ud}{\ud t} \int_\mathcal{V} N\, \sqrt{\gamma}\,\ud^3 x 
& = - \int_\mathcal{V} \frac{\partial\sqrt{\gamma}\,N\,\dot{x}^i}{\partial x^i}\,\ud^3 x \\
& = - \oint_\mathcal{S} (N\,\dot{x}^i) \sqrt{\gamma}\,\ud S_i,
\end{split}
\label{e1.14}
\end{equation} 
where $\mathcal{N}$ is the total number of baryons in the volume $\mathcal{V}$, $\mathcal{S}$ is the two-dimensional surface delimiting $\mathcal{V}$, and $\sqrt{\gamma}\,\ud S_i$ is a differential area element normal to $\mathcal{S}$. For the last step we have used the divergence theorem. If we take now the volume element $\mathcal{V}$ in \eq{e1.14} to coincide with the volume around a given SPH particle, we see that, by keeping a constant number of baryons within each individual SPH particle, both sides of \eq{e1.14} are equal to zero and the continuity condition in \eq{e1.12} is thus automatically fulfilled. 

\subsection{Self-gravity treatment}\label{sec:self-gravity}

In many situations, the BH dominates the spacetime, but the effects of self-gravity in the fluid cannot be entirely neglected.
For our purposes, we use a standard kernel-softened, Newtonian self-gravitational acceleration $\vec{a}_{\rm sg}$ and potential $\Phi$ calculated via a binary tree \citep{benz90b}. 
In the spirit of the Newtonian Euler equation, we introduce the force due to the fluid's self-gravity into the evolution equations by modifying \eq{e1.8} as
\begin{equation}
\begin{split}
\frac{\ud^2 x^i}{\ud t^2} = & -\left(g^{i \lambda}-  \dot{x}^i g^{0 \lambda}\right)
\bigg[ \frac{1}{\Gamma^2\omega}\left(\frac{1}{\varrho}\frac{\partial P}{\partial x^\lambda}  
+ \frac{\partial \Phi}{\partial x^\lambda} \right) + \\
& \hspace{40pt}\left( \frac{\partial g_{\mu\lambda}}{\partial x^\sigma} - \frac{1}{2}\frac{\partial g_{\mu\sigma}}{\partial x^\lambda} \right)\dot{x}^\mu\, \dot{x}^\sigma \bigg].
\end{split}
\label{e2.3}
\end{equation}

Since the way in which we treat self-gravity depends on the particular choice of spatial coordinates -- and hence is at odds with general relativity's covariance principle --, we test the validity of our approximate approach by comparing the outcome of the same simulation using two different coordinate systems (namely, Boyer--Lindquist (BL) and Kerr--Schild (KS) coordinates; see Sec.~\ref{sec:BLvsKS} below). Explicit calculations of all of the relevant terms in each coordinate system are given in Appendices \ref{AA} and \ref{AB}.

It is worth noting in \eq{e2.3} that, since self-gravity and pressure forces always enter the evolution equation together, hydrostatic equilibrium will be guaranteed as long as the two forces are comparable, and in the regime in which both are much larger than the tidal forces due to the BH. The prefactor $1/\Gamma^2$ effectively introduces a time dilation effect close to the BH (which is physical in the chosen coordinate system, i.e.~KS or BL), while the $g^{i \lambda}-  \dot{x}^i g^{0 \lambda}$ term introduces nonlinearity in the equations (e.g., the $x$ component of the acceleration will depend on all derivatives $\partial_\lambda P$ and $\partial_\lambda \Phi$, not just on $\partial_x P$ and $\partial_x \Phi$).

The evolution equation \eqref{e2.3} can be conveniently implemented within an existing SPH code with essentially no additional overhead in computational time with respect to a purely Newtonian calculation. 
To check this conclusion, we performed statistical timing measurements of both the relativistic and the Newtonian versions of our code, using $2\times 10^5$ and $10^6$ SPH particles, with both the serial (1 CPU core) and OpenMP (8 CPU cores) versions. On average, between 85 and 95 per cent of a time step is spent on tree operations (tree build, neighbour search, and Newtonian self-gravity calculations), while the rest of the time is mostly spent computing hydrodynamical and BH accelerations. Since the vast majority of the computational power is employed on parts of the code that are identical in both the Newtonian and the relativistic versions, the overhead for computing the relativistic corrections is between 0.5 and 2 per cent of a time step (depending on the number of particles and the number of CPU cores), the vast majority (about 80 per cent) of which is spent evaluating metric derivatives. 

Relativistic simulations will also take longer because the star spends more time close to periapsis as compared to the Newtonian simulations (due to the gravitational time dilation), but this is a physical effect that occurs in any relativistic simulation regardless of the code used, and therefore we do not consider it as an overhead of the method itself. Finally, we observed that during the periapsis passage, when the time step criterion based on the acceleration is more restrictive than the CFL criterion, the time steps taken in the relativistic simulations by our adaptive time-stepping scheme are, on average, smaller by a factor of $\simeq 2$ than those taken in Newtonian simulations. It is difficult to quantify the resulting computational overhead for an entire simulation, as the fraction of time spent close to periapsis depends considerably on the orbital parameters. Based on the simulations we ran for this paper, we estimate that this effect increases the computational cost by at most $\simeq 10$ per cent.

\setcounter{equation}{0}
\section{Validation}
\label{sec:validation}

In this section we present several tests designed to explore the range of applicability of our method. We shall do this by first testing the ability of the new approach to reproduce exact geodesic motion of test particles in both Schwarzschild and Kerr spacetimes. Next, we compare the outcome of several TDE simulations against the results of previous relativistic studies. Finally, we test the degree to which our method abides by the covariance principle by comparing the outcome of two simulations of the same TDE encounter performed once in Boyer--Lindquist and once in Kerr--Schild coordinates.

All simulations presented in this paper were conducted with a modified version of the Newtonian SPH code described in detail in \citet{rosswog08b}, where the Newtonian accelerations due to the BH and the pressure forces were replaced by the relativistic expressions given in Appendices \ref{AA} and \ref{AB}. The initial profile of the star was determined by solving the Lane--Emden equations for a $\gamma=5/3$ polytrope; the SPH particles (200\,642 in all cases) were distributed according to the resulting density profile, then relaxed with damping into numerical equilibrium (see~\citealp{rosswog09} for more details), and subsequently placed at a distance of $5\,\rt$ from the BH, on a parabolic orbit computed according to the equations given in Appendix~\ref{A2}. Throughout the simulation, the stellar fluid was evolved with a $\gamma=5/3$ polytropic equation of state. The BH accretion radius was placed at the event horizon radius $r_\mathrm{eh}$, see Eq.~\eqref{reh}, with all of the particles that entered this radius being removed at the next synchronization point of our individual time step scheme. Owing to the large mass ratio between the central BH and the star ($10^6$), the contribution of the accreted particles to the mass and spin of the BH has been neglected, i.e.~the metric is taken to be stationary. We then assume a fixed Kerr metric for a BH of mass $\mbh$, specific angular momentum $a$, and dimensionless spin parameter  
\begin{equation}
\abh \equiv \frac{a}{M}.
\end{equation}

We would like to stress at this point that in all of the numerical tests presented below we used the Euclidean distance for the calculation of all inter-particle separations. In an SPH code this distance is critical for building the tree itself (which is then used for computing the hydrodynamic and self-gravity accelerations), and then appears in all of the expressions that contain the SPH kernel or its derivatives, such as those for: gas density, momentum equation, energy equation, artificial viscosity terms, shock heating terms, and self-gravity acceleration. One could, in principle, also calculate the inter-particle separation via the proper spatial distance using the spatial metric tensor $\gamma_{ij}$ (rather than the Euclidean, flat-space distance that we are using), but this would only be an extra layer of complexity on top of an already approximate way of introducing Newtonian gravity within a relativistic approach and is therefore not considered.

The rationale behind our choice is that, since the BH mass is being taken to dominate the spacetime geometry i.e.~$M\gg \mst$, the spacetime will be very nearly flat across the scales where self-gravity matters (e.g., the radius of a star). On the other hand, on scales where the metric does significantly depart from flat-spacetime (two fluid elements on different sides of the BH being the most extreme example), self-gravity will have absolutely no relevant contribution.

A quantitative criterion for the previous argument can be stated by requiring the radius of curvature $\mathcal{R}$ of the BH's spacetime to be much larger than the length scales for both the self-gravity and the hydrodynamic interactions (that we take to be the radius of the star $\rst$ and the typical size for the SPH smoothing length $h$, respectively). The spacetime curvature around a BH is of the order of $\kappa = M/r^3$, from where it follows a radius of curvature $\mathcal{R}\propto 1/\sqrt{\kappa} = 1.5 (M/\msun) (r/\rg)^{3/2}\,$km. From here we see that, in the encounter between a solar-type star ($\rst\simeq 7\times10^5\,$km) and a $10^6\,\msun$ BH, $\mathcal{R}/\rst \simeq 700$ at a distance of one tidal radius from the BH (from Eq.\eqref{e0.1}, $r_t\simeq 48\,\rg$ in this case). This same ratio reduces to about 30 at a periapsis distance of $\rp  = 6\,\rg$. On the other hand, for the simulations presented in this article, a typical smoothing length used in the SPH description of the fluid will be at least one order of magnitude less than the corresponding stellar radius. Clearly, the ratio $\mathcal{R}/\rst$  evaluated at a fix number of gravitational radii from the BH, say at $ 6\,\rg$, becomes larger (more favourable) for more massive BHs.\footnote{Note however that, for a fixed stellar type, the ratio $\mathcal{R}/\rst$ evaluated at the tidal radius acquires a constant value independent of the BH mass.} This gives us confidence that our approximation should be valid for the type of TDEs discussed in the next sections.

\begin{figure*}
\centerline{
\includegraphics[height=.46\linewidth]{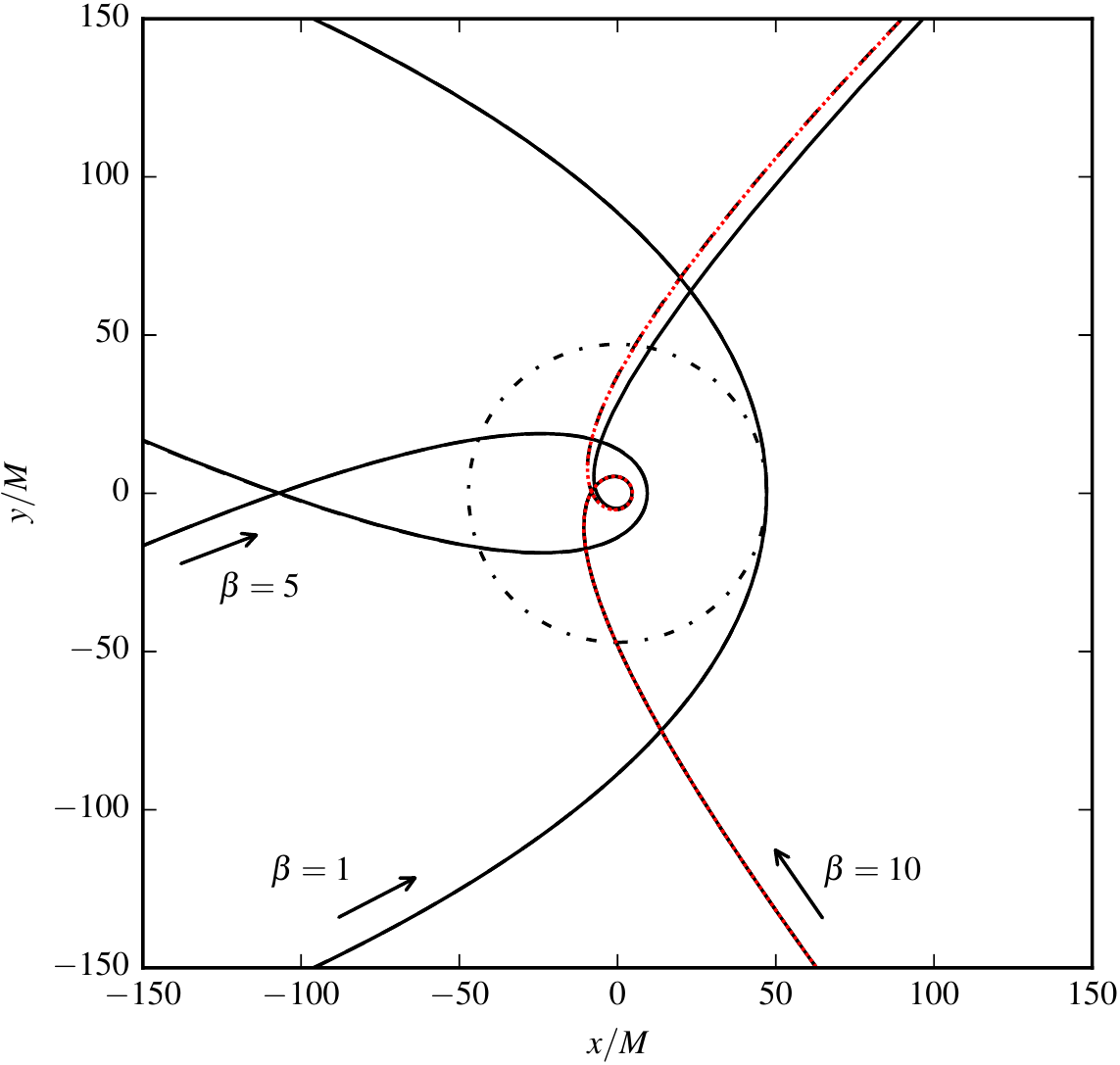}\hfill
\includegraphics[height=.46\linewidth]{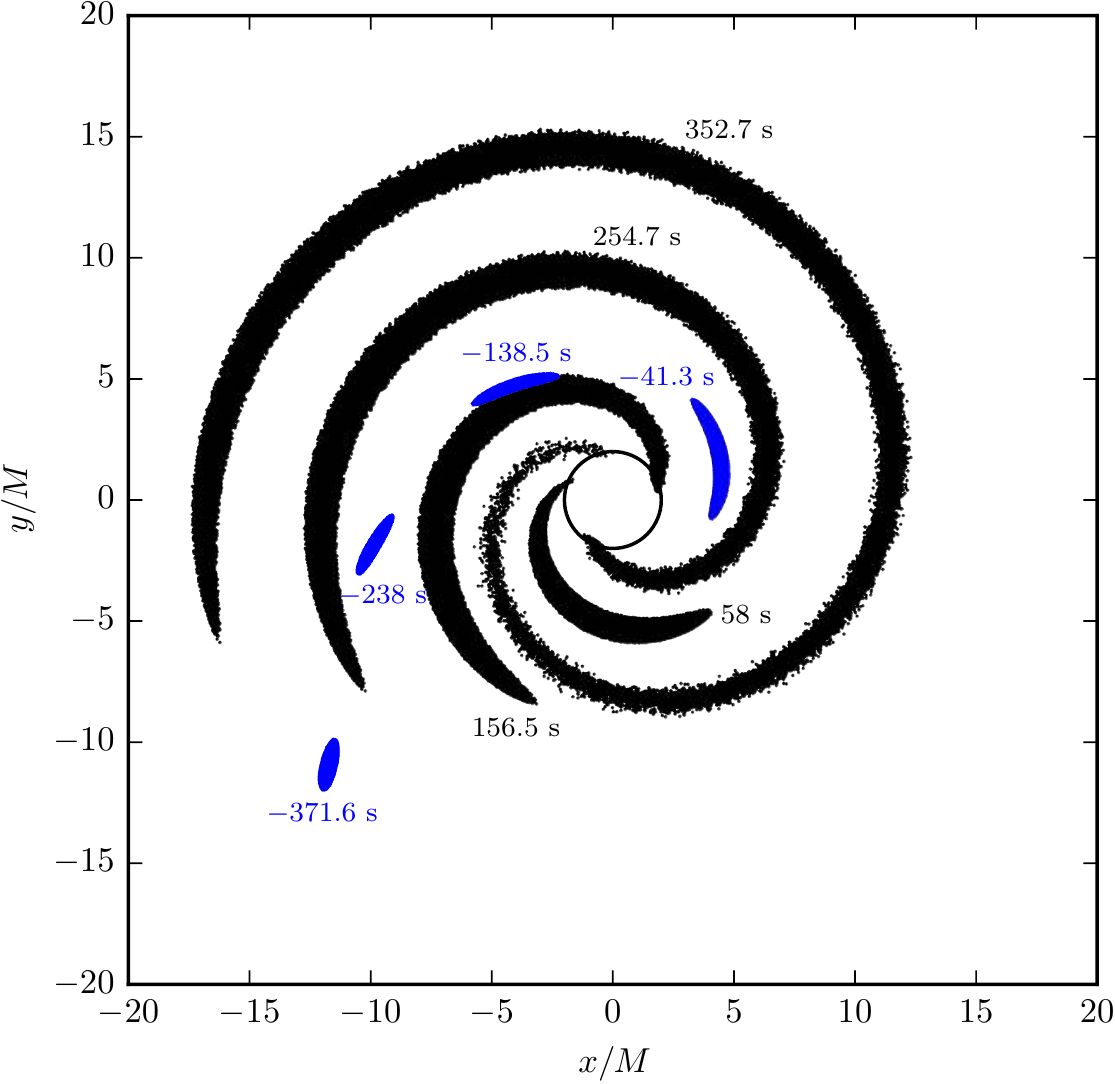}}
\caption{Tidal disruption of a solar-type star by a $10^6~\msun$ Schwarzschild BH.
\textit{Left panel.} Trajectory of the centre of mass of the star (solid line), compared with the geodesic trajectory (dashed line), for
three canonical TDEs with impact parameters $\beta=1$, 5, and 10, respectively.  The tidal radius of the BH is shown as a dash-dotted circle. For $\beta=1$ and 5 our results are indistinguishable from the exact geodesic motion (and hence the dashed line is not clearly visible), but for the $\beta=10$ case some deviations are expected, see main text, Sec.~\ref{sec:geomlim}. 
For this reason we also plot the trajectory of the centre of mass of the core 200 particles (dotted red line), which is indistinguishable from the geodesic (see main text, Sec.~\ref{sec: validation-geodesic}, for justification and interpretation).
\textit{Right panel.} Snapshots of the disrupted star with $\beta=10$, at various times before the disruption (blue) and after the disruption (black) ($t=-371.6, -238, -138.5, -41.3; 58, 156.4, 254.7, 352.7$ s). }
\label{fig:CMlaguna}
\end{figure*}

\subsection{Geodesic motion limit}\label{sec:geomlim}

For an ensemble of pressureless, non-interacting particles, the evolution equation given in \eq{e1.8} naturally reduces (by setting $P=0$) to the geodesic equation, i.e.
\begin{equation}
\frac{\ud^2 x^i}{\ud t^2} = -\left(g^{i \lambda}-  \dot{x}^i g^{0 \lambda}\right)
\left( \frac{\partial g_{\mu\lambda}}{\partial x^\sigma} - \frac{1}{2}\frac{\partial g_{\mu\sigma}}{\partial x^\lambda} \right)\dot{x}^\mu\, \dot{x}^\sigma ,
\label{e3.1}
\end{equation}
which is more commonly expressed in terms of proper time derivatives as \citep*{mtw}
\begin{equation}
\frac{\ud^2 x^\mu}{\ud \tau^2} + \Gamma^\mu_{\nu\lambda}\, \frac{\ud x^\nu}{\ud \tau} \,
 \frac{\ud x^\lambda}{\ud \tau} = 0 .
 \label{e3.2}
\end{equation}

The geodesic equations \eqref{e3.1} constitute a system of second order, coupled ordinary differential equations. Nevertheless,  in the case of the Kerr spacetime, the existence of four first integrals of motion allows us to partially decouple and reduce them to a set of first-order ordinary differential equations which can be further solved analytically \citep[e.g.][]{bardeen73,chandra,novikov,tejeda3}. See Appendix \ref{A} for a brief overview of time-like geodesics in this spacetime. In this appendix we also outline the procedure which we follow to generate initial conditions from a given set of orbital parameters.

Throughout the rest of the paper, all of the geodesic trajectories that we present (for instance, in comparison with the orbits of the centres of mass) result from the direct integration of Eq.~\eqref{e3.1} for a point mass (with the same initial constants of motion as the stellar centre of mass) using a fourth-order Runge--Kutta integrator (RK4), completely independent of our SPH code.

In a TDE, \eq{e3.1} will be obeyed by the trajectory of the star's centre of mass (CM) (with some caveats discussed below), and by the individual trajectories of the fluid elements after the star has been disrupted, when self-gravity and hydrodynamic forces no longer play a significant role. A common way to evaluate how well an approximate method reproduces effects such as periapsis shift and orbital plane precession is to compare the trajectory of the CM with the corresponding geodesic (i.e., the one having the same constants of motion). While this is in many cases a meaningful comparison, there are cases where deviations are expected.

\begin{figure*}
\centering
\includegraphics[width=\linewidth]{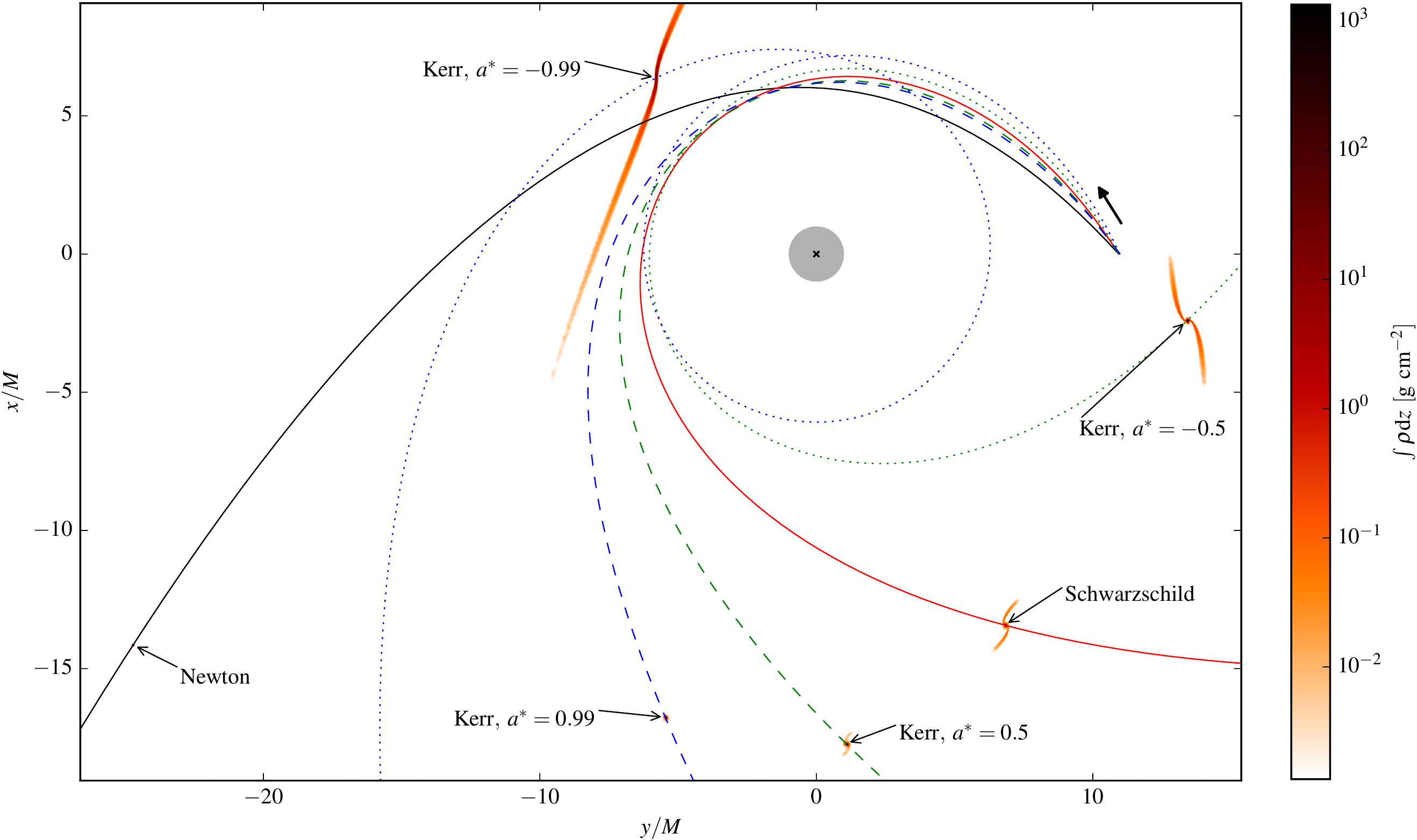}
\caption{
Tidal disruption of a solar-type star ($\mst=\msun$, $\rst=\rsun$) by a $10^8~\msun$ Kerr BH for different values of the spin parameter. The snapshots show the matter distribution $\approx 12$ hours after the first periapsis passage, while the lines represent the geodesics of the corresponding orbits. The periapsis distance in all cases is $ \rp=6\,\mbh$, corresponding to an impact parameter $\beta\simeq 0.36$. The solid black curve represents the Newtonian simulation; the solid red curve represents the disruption by a non-rotating ($\abh = 0$) BH; the dashed (dotted) curves represent the trajectories of stars on prograde (retrograde) orbits around BHs with spin parameters 0.5 (green) and 0.99 (blue). The simulations were made using KS coordinates. For clarity purposes, the stellar debris has been \textbf{magnified by a factor of 10} for all six simulations; the magnification is performed as a geometric scaling operation during rendering, but the colours are preserved in order to yield the correct density.
 The BH position is marked by a black cross, while the gravitational radius $\rg$ is marked by a grey disc.}
\label{fig:KerrSpinsDens}
\end{figure*} 
 
First, for deep encounters (impact parameter $\beta\gg 1$), where complete disruption occurs from the very first periapsis passage, the wide spread in specific orbital energies and accompanying redistribution of angular momentum results in a fraction of the fluid being launched on plunging orbits. As such fluid elements are accreted, the CM of the debris stream is effectively calculated on a different set of particles and therefore the CM trajectory deviates from the original geodesic. An example of such a case (for $\beta=10$) can be seen in the right-hand panel of Fig.~\ref{fig:CMlaguna}, which shows the disruption of a solar-type star by a $10^6~\msun$ Schwarzschild BH, which will be discussed further below.

Another effect present in deep encounters is related to the geometrical shape of the debris. Typically, once a star is disrupted and the fluid elements move along (nearly) independent geodesics, each having experienced a different periapsis shift, the stellar debris will expand into a crescent shape (first observed and discussed by \citealp{laguna93b}). For encounters with small $\rp/M$ ratios, the large spread in periapsis shifts may stretch the crescent-shaped debris into a spiral (as seen before in the simulations of \citealp{kobayashi04} and \citealp{cheng14b}; see also Fig.~\ref{fig:CMlaguna}), causing the CM to drift outside the particle distribution; as the physical extent of the debris becomes comparable to its distance from the BH, the CM is no longer meaningful in providing information about the motion of the fluid as a whole around the BH. This effect is only relevant for relativistic simulations, where the stream is deformed by periapsis and orbital plane precessions.

Finally, a departure from geodesic motion also occurs due to the fact that orbital energy is converted into heat deposited in the oscillation modes of the star, which changes the total energy of the fluid, even though the dissipated energy ($\propto \mst/\rst$) is normally a few orders of magnitude smaller than the typical spread in energies at periapsis ($\propto M \rst/ {\rp}^2$), and therefore does not produce noticeable effects.

\subsubsection{Schwarzschild spacetime}\label{sec: validation-geodesic}

The simplest possible test is for a TDE involving a non-rotating BH: in such a case, the orbital motion is confined to one plane, and the main qualitative relativistic effect is the periapsis precession. 
The left-hand panel of Fig.~\ref{fig:CMlaguna} shows the trajectory of the CM (solid lines) and the geodesic trajectory (dashed lines) for three simulations of canonical ($\mst=\msun$, $\rst=\rsun$, $\mbh=10^6~\msun$, $\abh=0$) TDEs with impact parameters $\beta=1$, 5 and 10. These results can be directly compared with those obtained using relativistic SPH codes by \citet[][Fig.~1]{laguna93b} and \citet[][Fig.~2]{kobayashi04}. In agreement with their results, we find the trajectory of the  CM  to be indistinguishable from the geodesic for $\beta=1$ and 5, while the small departure that occurs for $\beta=10$ is caused by a combination  of the reasons discussed above. 

As support for our interpretation of this departure, for the $\beta=10$ case we also plot the trajectory of the centre of mass for the $N_{\rm c}$ core SPH particles, i.e.~the $N_{\rm c}$ particles that -- at the beginning of the simulation -- are the closest to the centre of mass of the star. The number $N_{\rm c}=200$ has been chosen so that, on the one hand these particles are neither accreted, nor does their centre of mass drift away from the gas distribution, and, on the other hand the discretization errors are small enough for their CM to reproduce the geodesic. For a much smaller $N_{\rm c}$, the average energy and angular momentum would be slightly different from those of the star and of the geodesic, failing to reproduce geodesic motion; for a much larger $N_{\rm c}$, the geometric distortion would produce the same departure from geodesics that occurs for the CM of the star itself, as explained above.

The right-hand panel shows snapshots of the tidal debris before, during, and after the first periapsis passage in the $\beta=10$ disruption; our results are in excellent agreement with the simulation of \citet[][Fig.~3]{kobayashi04}. 

\subsubsection{Kerr spacetime, equatorial orbits}

Next, we simulated a TDE involving a rotating BH in the particular case in which the star's CM motion is confined to the equatorial plane (i.e., to the plane perpendicular to the BH spin). 
Previous simulations by e.g.~\citet{haas12} and \citet{kesden12} found that, for a given impact parameter, stars are more easily disrupted if they approach the BH on a retrograde orbit (represented here as a negative BH spin parameter $\abh<0$, cf.~Section \ref{sec:rel_effects}). Fig.~\ref{fig:KerrSpinsDens} shows the distribution of the stellar debris after disruption by a BH with $\mbh=10^8~\msun$ and various spin parameters ($\abh=0$, $\pm 0.5$, and $\pm 0.99$). We simulated the disruption of a solar-type star ($\mst=\msun$, $\rst=\rsun$) on an orbit with periapsis distance $ \rp=6\,\mbh$.
Since this corresponds to an impact parameter of $\beta\simeq 0.36$, the Newtonian simulation does not result in disruption: the star is barely tidally deformed, and simply continues along its original trajectory. The relativistic effects at $6\,\mbh$, however, are strong enough that all of the Kerr simulations (and the Schwarzschild one) result in strong tidal deformation, the formation of tidal tails, and (with the exception of the $\abh=0.99$ simulation) some fraction of the stellar material being stripped away. This fraction of unbound material is given by $f_{\rm ub}=1-f_{\rm b}$, where the self-bound fraction $f_{\rm b}$ is computed using the iterative-based prescription introduced by \citet[][Sec. 2.2]{guillochon13}, which we have previously used successfully with SPH in \citet{GT15}. During this iterative procedure, the (Newtonian) gravitational self-potential of the stellar debris was computed using a fast binary tree \citep{gafton11}.

The fraction $f_{\rm ub}$ depends strongly on the BH spin, and spans from zero (for $\abh=0.99$), to less than one per cent (for $\abh=0$), to over 60 per cent (for $\abh=-0.99$). This result is in agreement with the observation of \citet{GT15}, that relativistic effects connected to the amplified tidal stresses and the extra time spent close to periapsis result in stronger disruptions (even for relatively low $\beta$). Note, for instance, that for $\abh=-0.99$ the periapsis shift is so strong that the star completes one full winding around the BH before receding.

\begin{figure}
\centering
\includegraphics[width=\linewidth]{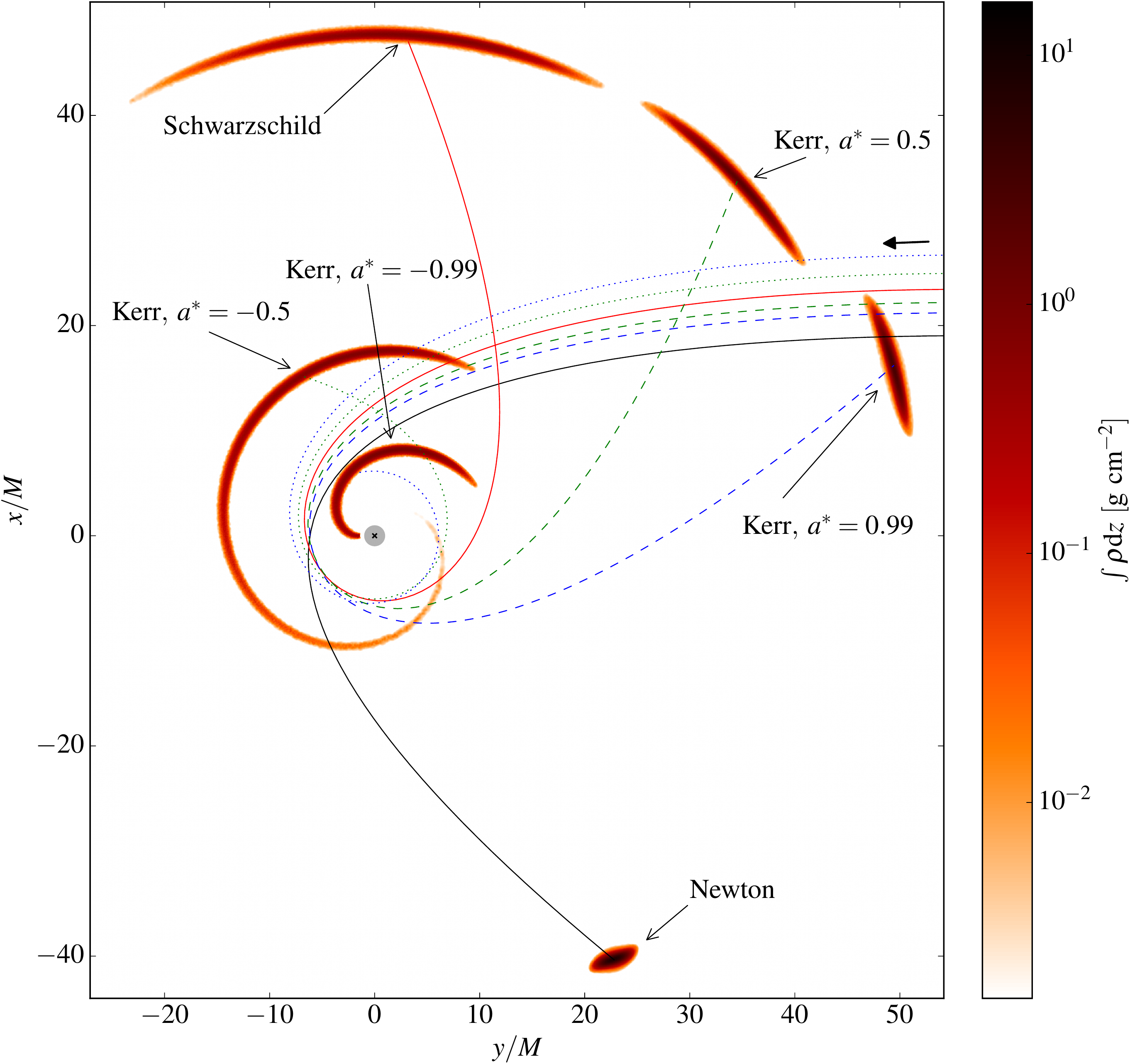}
\caption{Tidal disruption of a solar-type star ($\mst=\msun$, $\rst=\rsun$) by a $10^6~\msun$ Kerr BH for different values of the spin parameter. The snapshots show the matter distribution $\approx 15$ minutes after the first periapsis passage, while the lines represent the geodesics of the corresponding orbits. The periapsis distance in all cases is $\rp=6\,\mbh$, corresponding to an impact parameter $\beta\simeq 7.85$. The solid black curve represents the Newtonian simulation; the solid red curve represents the disruption by a non-rotating ($\abh = 0$) BH; the dashed (dotted) curves represent the trajectories of stars on prograde (retrograde) orbits around BHs with spin parameters 0.5 (green) and 0.99 (blue). The simulations were made using KS coordinates. The gravitational radius $\rg$ is marked by a grey disc.}
\label{fig:CMKesden}
\end{figure}

We also ran a similar set of simulations for $\mbh=10^6~\msun$. In this case, $\rp=6\,\mbh$ corresponds to an impact parameter of $\beta\simeq 7.85$, and so all of the simulations resulted in the star being fully disrupted. Fig.~\ref{fig:CMKesden} shows the distribution of the stellar debris after disruption, together with the various geodesic trajectories corresponding to the CM. All of the relativistic simulations resulted in a strong disruption as compared with the Newtonian simulation: the maximum density is lower and the debris stream is more elongated. For the retrograde orbits, $6\,\mbh$ is very close to the marginally bound circular orbit (located at $\approx 5.8\,M$), and therefore some particles are launched on plunging orbits from the very first periapsis passage. This leads to prompt accretion and a spiral-shaped debris stream around the BH. For prograde orbits, disruptions are increasingly milder with increasing BH spin, though always stronger than their Newtonian counterpart. 

\subsubsection{Kerr spacetime, off-equatorial orbits} 

\begin{figure*}
\centering
\includegraphics[width=\linewidth]{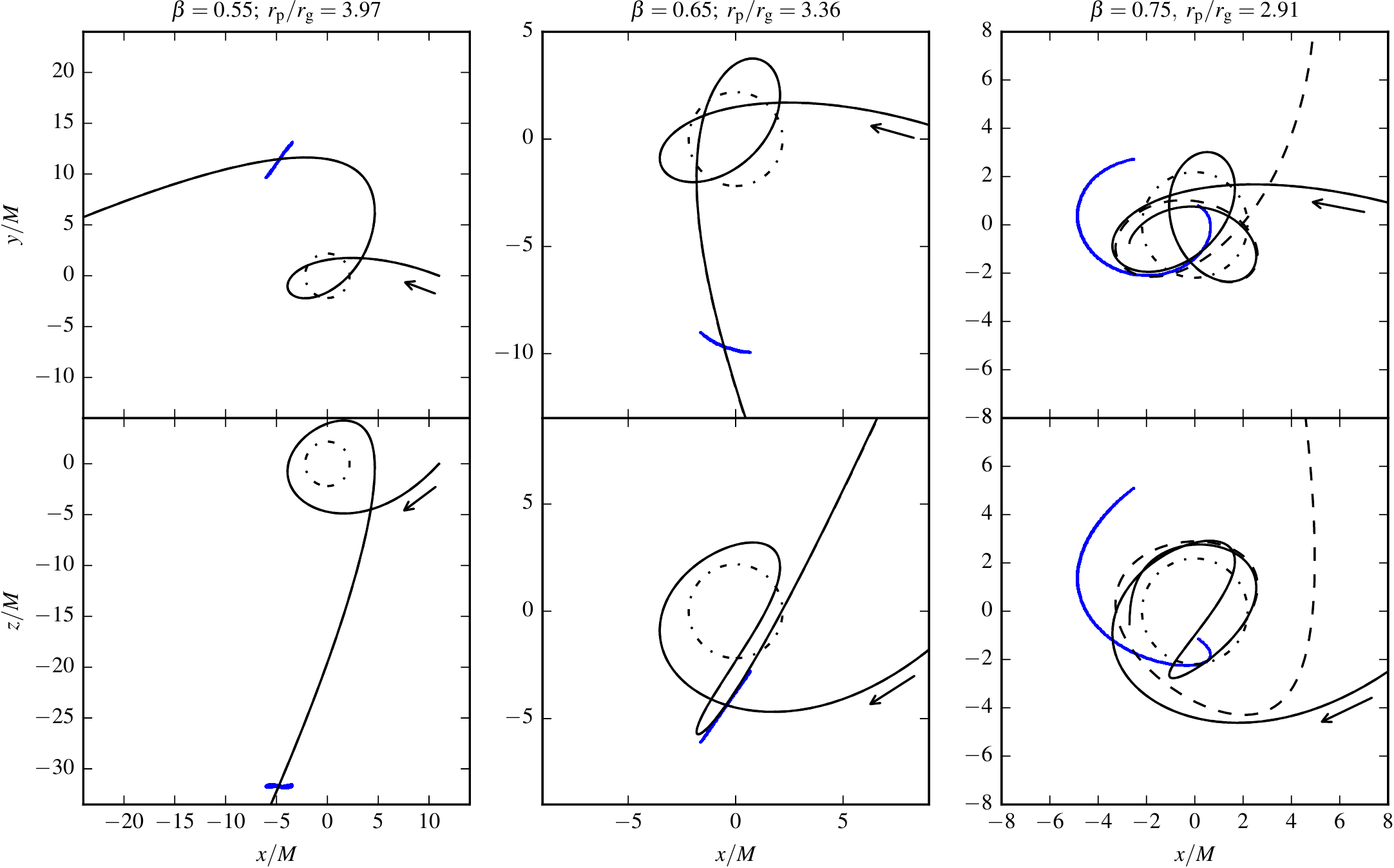}
\caption{Trajectory of the CM of the star (solid line), compared with the geodesic motion (dashed line) for three TDEs around a Kerr BH (of mass $\mbh=10^8~\msun$ and spin parameter $\abh=0.98$) with three different values of the impact parameter ($\beta=0.55$, 0.65, and 0.75). Note that in the first two cases the dashed and solid lines are indistinguishable from each other. The top and bottom panels show projections onto the $x-y$ and $x-z$ planes, respectively, with the dash-dotted circles marking the tidal radius.
The blue scatter plots show the SPH particles at times $t \simeq 21$, 7.5, and 5 h  (for $\beta=0.55$, 0.65, and 0.75, respectively) after the disruption, as they recede from the central BH. The CM trajectory in the last case ($\beta=0.75$) has been interrupted at the point where the tidal stream becomes so elongated, with a size comparable to the distance from the BH, that the CM is no longer meaningful (see Sec.~\ref{sec:geomlim}).}
\label{fig:CMKerr}
\end{figure*}

\begin{figure*}
\centering
\includegraphics[width=0.9\textwidth]{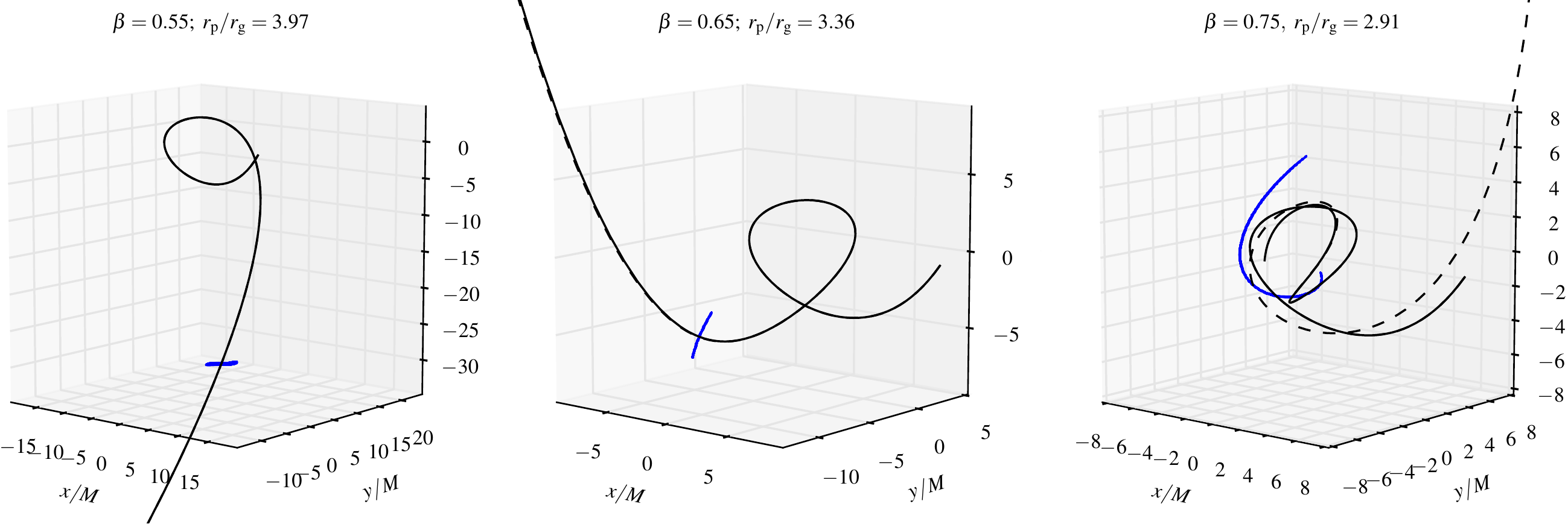}
\caption{Three-dimensional version of Fig.~\ref{fig:CMKerr}; again, the trajectory of the stellar centre of mass (solid line) is compared with the geodesic trajectory (dashed line); the three-dimensional particle distribution after disruption is shown as a blue scatter plot.}
\label{fig:CMKerr3D}
\end{figure*}

Finally, we simulated a disruption of a solar-type star approaching a rotating BH ($a^*=0.98$) along an off-equatorial trajectory. The star was initially located in the equatorial plane but had a non-zero polar angular velocity. This results in an angular span for its latitudinal motion ranging from a minimum latitude ($\theta_a=0.1\,\pi$) to a maximum one ($\theta_{a'}=0.9\,\pi$) (see Appendix~\ref{A2} for details). In this situation, orbital plane precession is expected to play a significant role in the shape and evolution of the stellar debris.
Fig.~\ref{fig:CMKerr} shows the trajectory of the CM (solid line) and the geodesic trajectory (dashed line) for three TDEs with increasing impact parameters ($\beta=0.55$, 0.65, and 0.75) around a Kerr BH with $\mbh=10^8~\msun$ and spin parameter $\abh=0.98$. The upper (lower) panels show the projection of the orbit onto the $x-y$ ($x-z$) plane. We also show a scatter plot of the SPH particles in blue. Fig.~\ref{fig:CMKerr3D} shows a three-dimensional view of these same three encounters.

In all three simulations, relativistic effects are reproduced very well: for $\beta=0.55$ and $0.65$, there is virtually no  difference between the CM and the geodesic trajectories, while for $\beta=0.75$, after several windings around the BH, the stream becomes so distorted that the CM drifts off the geodesic trajectory. As discussed above, this is simply a geometric effect. The orbits of the individual particles, which at this time are moving as point masses in the gravitational field of the BH, still follow their respective geodesic trajectories.

\subsection{Comparison between Boyer--Lindquist and Kerr--Schild coordinates}
\label{sec:BLvsKS}

In this section we explore the validity of our approximate approach by comparing the output of simulations of the same tidal disruption event as computed using two different coordinate systems: Boyer--Lindquist (BL) and Kerr--Schild (KS) coordinates. Useful expressions for both coordinate systems are collected in Appendices \ref{AA} and \ref{AB}.

Due to the principle of covariance, two relativistic simulations performed in different coordinate systems should give identical physical results. Since geodesic motion in Kerr spacetime is reproduced exactly with our code, any difference between the BL and KS simulations will come from the Newtonian parts of the code: the inclusion of Newtonian self-gravity (albeit with relativistic corrections as described in Sec.~\ref{sec:self-gravity}), and the calculation of inter-particle distances (without the use of the metric), which enters into the expressions for all inter-particle forces.

\begin{figure*}
\includegraphics[width=.42\linewidth]{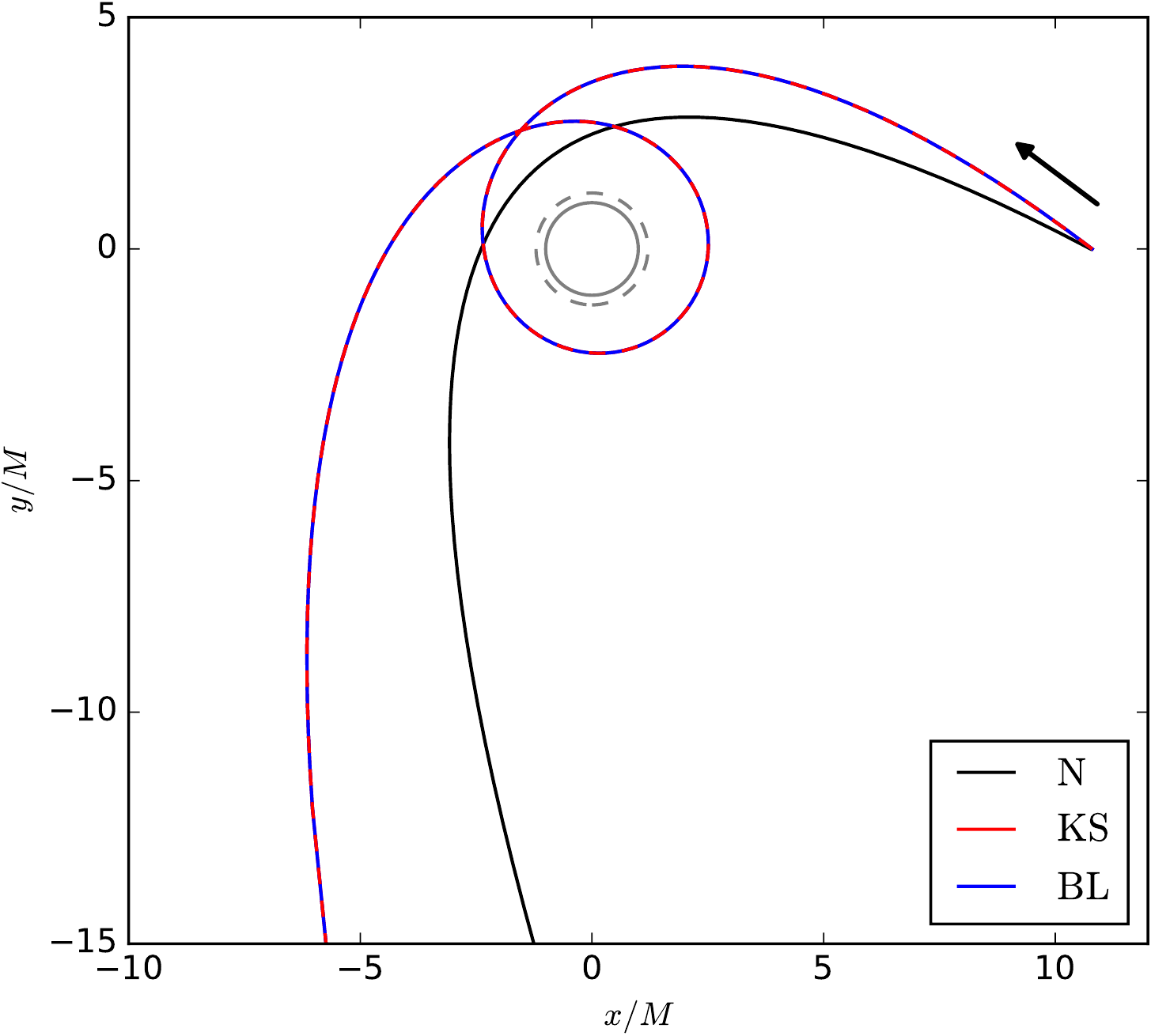} 
\includegraphics[width=.57\linewidth]{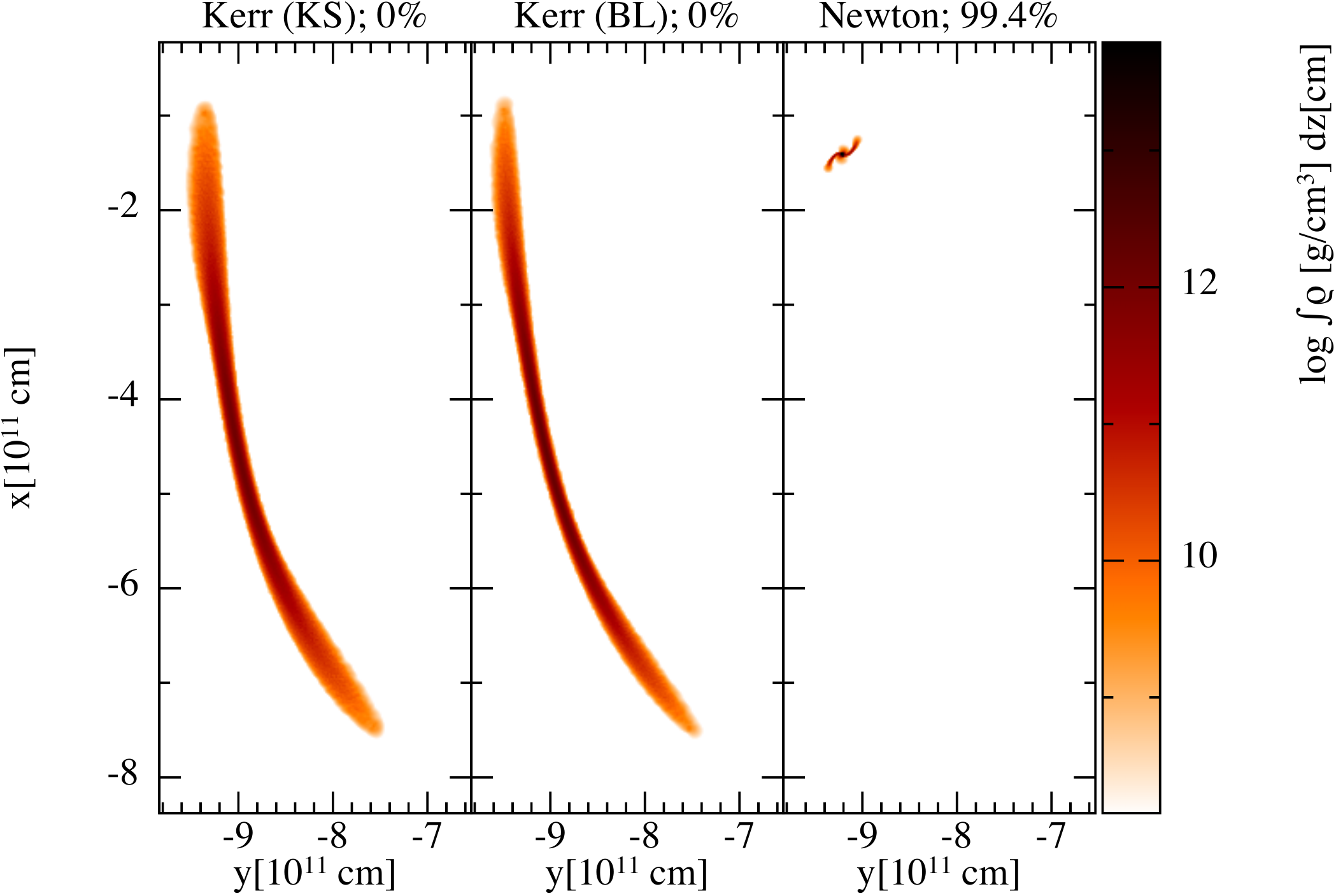}
\caption{Comparison between three simulations with the same initial conditions ($\mbh=5\times 10^5~\msun$; $\abh=0.98$; $\mst=0.6\,\msun$; $\rst=0.014\,\rsun$; $\beta=0.55$): relativistic Boyer--Lindquist (BL), relativistic Kerr--Schild (KS), and Newtonian (N). \textit{Left panel.} The trajectory of the centre of mass (solid lines) compared with the geodesics (dashed lines). The BH gravitational radius and the tidal radius are marked as solid and dashed grey circles, respectively. \textit{Right panel.} The spatial distribution of the debris after disruption, colour-coded by density integrated along the line-of-sight. The percentage of self-bound debris is given in the title of each panel: the encounter results in a complete disruption in both KS and BL coordinates, but strips away less than 1 per cent of the star in the Newtonian case. In both panels only KS coordinates are being used for plotting; results obtained in the BL simulation are trasformed into KS coordinates as part of the post-processing before plotting.}
\label{fig:KSvsBLvsN}
\end{figure*}

In Fig.~\ref{fig:KSvsBLvsN} we show the CM trajectory (left panel) and a post-disruption snapshot of the stellar debris (right panel) for a highly relativistic ($\rp/\rg=2.19$), but mildly disruptive ($\beta=0.55$) simulation of the tidal disruption of a white dwarf by a $5\times 10^5~\msun$ rotating BH with spin parameter $\abh=0.98$. As expected for $\beta=0.55$, the Newtonian encounter only results in a negligible fraction ($0.6\%$) of the star being stripped away and forming two very weak tidal tails. On the other hand, the relativistic simulations result in a complete disruption of the star. From the left panel of this figure we see that the CM trajectories of the BL and KS simulations are indistinguishable from each other. Furthermore, both relativistic simulations result in full disruptions, and there is no qualitative difference in shape or density profile between the two snapshots of the debris stream (BL and KS), both taken at $t \simeq 2.5$ minutes after the first periapsis passage. 

\begin{figure*}
\centering
\includegraphics[width=\linewidth]{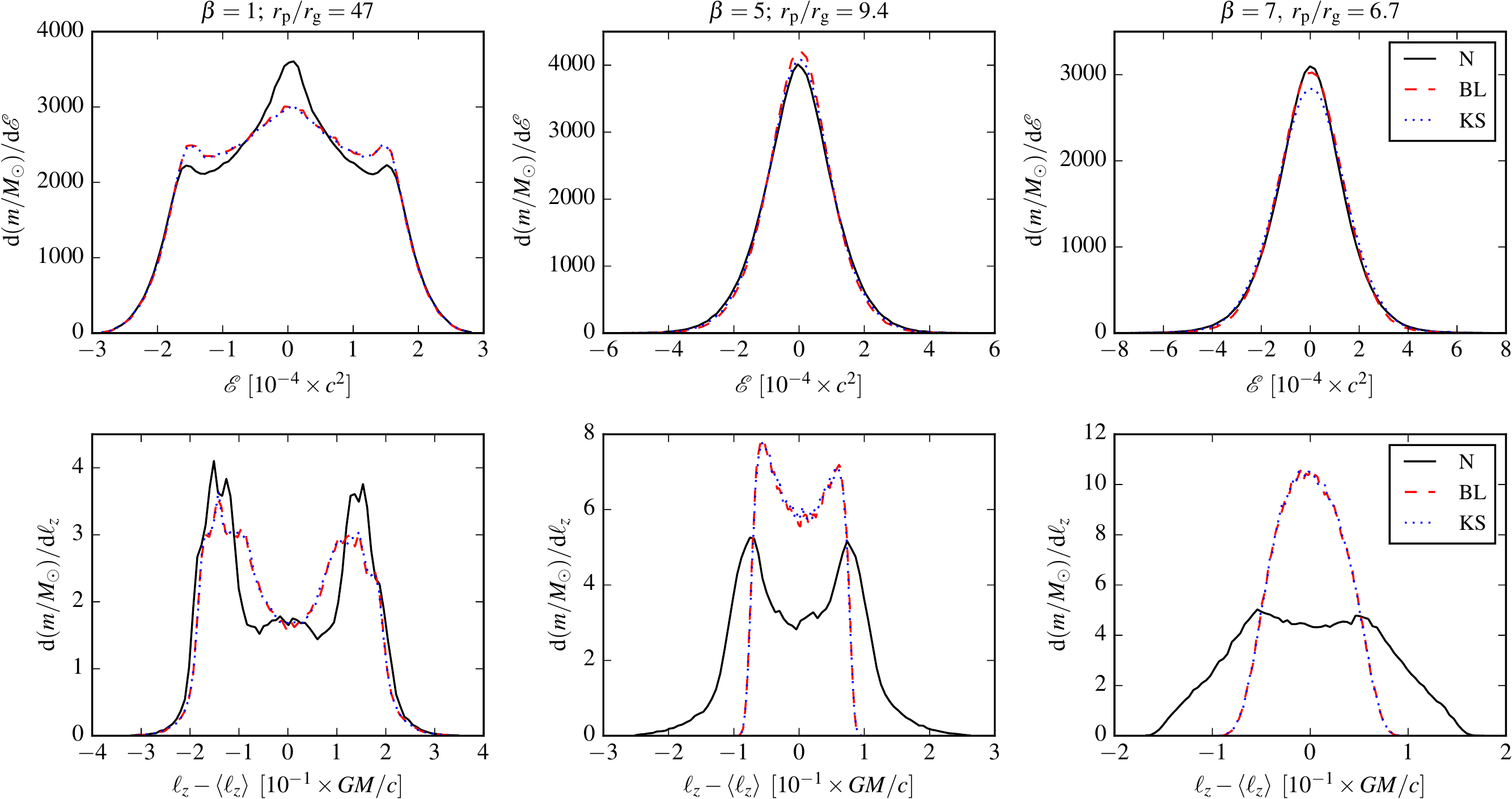}
\caption{Histograms of the two constants of motion relevant in a TDE around a non-rotating BH ($\abh=0$): the specific mechanical energy $\en$, and the specific angular momentum $\lz$; the Newtonian results (N; solid black line) are compared with the relativistic results computed in Boyer--Lindquist (BL; dashed red line) and Kerr--Schild (KS; dotted blue line) coordinates. We point out that while all energies are centred around $0$, the angular momentum in the Newtonian and relativistic simulations differs significantly (in order to get an orbit with the same $r_0$ and $\beta$), therefore we plotted the difference between $\lz$ and the average initial value $\left<\lz\right>$, which had the following values: for $\beta=1$, 9.70 for N and 9.92 for KS and BL; for $\beta=5$, 4.34 for N and 4.89 for KS and BL; for $\beta=7$, 3.67 for N and 4.37 for KS and BL. Even for strong encounters with $\beta=7$, the BL and KS results are in excellent agreement.
}
\label{fig:ELhistograms}
\end{figure*}

We also ran several simulations of canonical ($\mbh/\mst=10^6$) TDEs with $\beta=1$, 5, 7  in order to perform a quantitative comparison between  the results obtained in the two coordinate systems. In Fig.~\ref{fig:ELhistograms} we plot histograms of the constants of motion, i.e.~the energy $\en$ (top panels) and angular momentum $\lz$ (lower panels) for Newtonian (solid lines), BL (dashed line) and KS (dotted line) simulations, see Eqs.~\eqref{eq:A3} and \eqref{eq:A4} in Appendix~\ref{A}. The $\lz$ histograms are shifted and re-centred around zero, since the Newtonian values are strongly offset from the BL and KS ones. The agreement between KS and BL is excellent in spite of the Newtonian parts of the code.
The energy histograms exhibit differences of the order of 3 per cent for $\beta=5$ and 6 per cent for $\beta=7$, but the angular momentum histograms are virtually identical (within less than one percent) even at $\beta=7$.

In summary, we conclude that even for very deep encounters the results of the BL and KS simulations are in excellent agreement. This gives confidence that, at the very least, moderate (but possibly also strong) relativistic encounters can be accurately simulated  with our method.

\setcounter{equation}{0}
\section{Applications}
\label{sec:app}
In the previous section, we focused on the region near to the BH so as to make tests under significantly relativistic conditions. Now we turn to the more extended range as illustrated in Fig.~\ref{fig:disruption_geometry}.
In order to demonstrate some of the possible applications of our new method, we present here the results of three sets of TDE simulations for rotating BHs and analyse the impact of the BH spin on: a) the distribution of the constants of motion ($\en$ and $\lz$) after disruption, and hence on the mass fallback rate; b) the spread in mechanical energies $\en$ after disruption, as compared with Newtonian simulations and the analytical estimates of \mbox{\citet{stone13}}; c) the exotic geometry of the tidal debris in the case of one particular set of orbital parameters that has never been simulated before.

\subsection{Impact of spin on mass fallback rate}\label{sec:5.1}

We first simulated two sets of canonical TDEs ($\mbh=10^6~\msun$, $\mst=\msun$, $\rst=\rsun$) with impact parameters $\beta=2$ and $\beta=6$. In each case, we compared the results of a Newtonian simulation with five relativistic simulations, one without spin ($\abh=0$), and the other four with BH spins  $\abh=\pm 0.5$ and $\pm 0.99$, respectively. The $\beta=6$ is chosen as a limiting case, so that the $\abh= -0.99$ simulation (which yields the most extreme disruption) would not result in part of the debris being launched on plunging orbits from the very first periapsis passage. 

\begin{figure*}
\centering
\includegraphics[width=\linewidth]{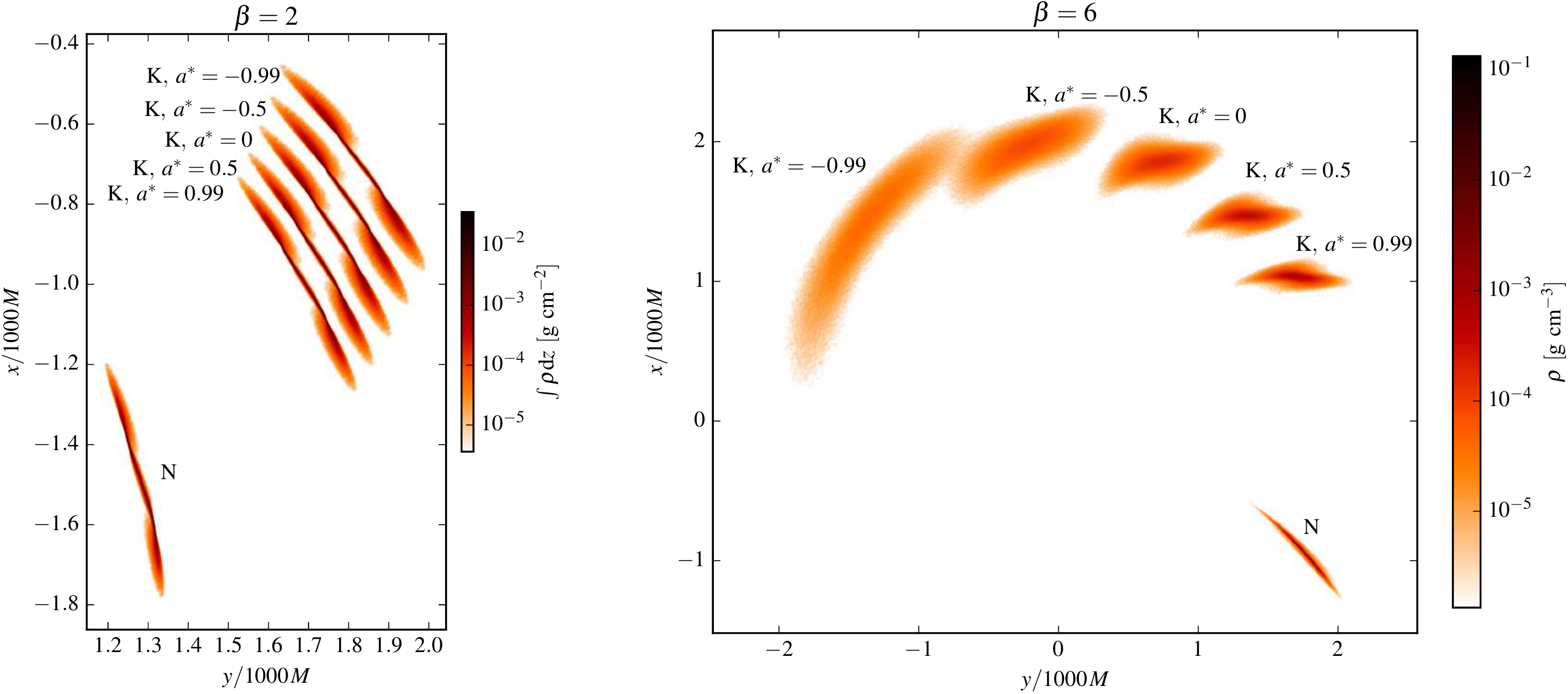}
\caption{Spatial distribution of the tidal debris for the canonical TDEs with $\beta=2$ (left panel) and $\beta=6$ (right panel) discussed in Sec.~\ref{sec:5.1}. For $\beta=2$, the effects of the BH spin are small: the matter distribution looks very similar at $t\simeq 57$~h after disruption, and is simply shifted due the different periapsis shifts. For $\beta=6$, however, the Newtonian encounter results in a markedly different debris distribution (a thin stream returning to the BH) from the relativistic simulations, which result in thick debris streams. The extent of the disruption (quantifiable, for instance, by the maximum density of the debris) and the thickness are larger for the Kerr simulations, and progressively increases as $\abh$ goes from $+0.99$ to $-0.99$.}
\label{fig:spinMdot2}
\end{figure*}

\begin{figure*}
\centering
\includegraphics[width=\linewidth]{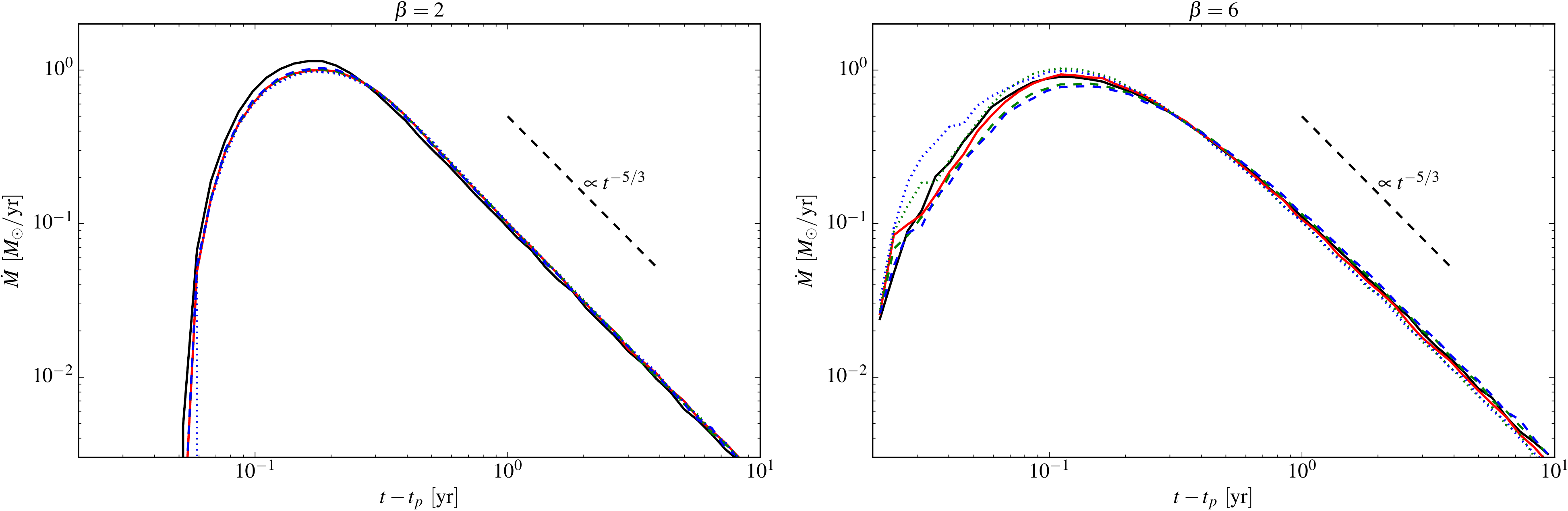}
\caption{Mass fallback rates $\dot{M}$ of the debris after the first periapsis passage for the canonical TDEs with $\beta=2$ (left panel) and $\beta=6$ (right panel) discussed in Sec.~\ref{sec:5.1}. The solid black curve represents the Newtonian simulation; the solid red curve represents the disruption by a non-rotating ($\abh = 0$) BH; the dashed (dotted) curves represent the trajectories of stars on prograde (retrograde) orbits around BHs with spin parameters 0.5 (green) and 0.99 (blue).}
\label{fig:spinMdot6}
\end{figure*}

The results are presented in Figures \ref{fig:spinMdot2} (geometric distribution of the tidal debris at $t\simeq 57$ hours after the first periapsis passage) and \ref{fig:spinMdot6} (mass fallback rates $\dot{M}$).
Each of these figures contains two panels, for $\beta=2$ (left) and for $\beta=6$ (right).
The mass fallback rate $\dot{M}$ is computed by assuming that particles move on geodesic trajectories after disruption (i.e., that hydrodynamic and self-gravity forces no longer play any role in their dynamics), and their trajectory is extrapolated based on their constants of motion. This yields a Keplerian orbit in the Newtonian case, but contains the proper relativistic corrections to the geodesics in the relativistic cases.

For $\beta=2$, there is only a minor difference in the debris distribution between the
Newtonian and relativistic simulations, see left panel of Fig.~\ref{fig:spinMdot2}. The BH spin does 
determine the position of the centre of mass at the given time, but the shape of the debris 
seems rather unaffected by the BH spin. In spite of the morphological
differences, the impact on the mass fallback rates is negligible, see Fig.~\ref{fig:spinMdot6}. 
All of the $\beta=2$ simulations result in very similar $\dot{M}$ curves. The relativistic ones are rising slightly more slowly (in about a week), partly due to the extra time spent around periapsis, though mostly due to the Newtonian energy spread being slightly larger (see Fig.~\ref{fig:espread} for $\beta=2$), thus resulting in more particles with lower eccentricities. Nonetheless, all of the relativistic simulations give essentially the same peak $\dot{M}$ rate, which is $\approx 13\%$ smaller than the Newtonian value; there is, however, no discernible influence from the BH spin.

For $\beta=6$, the differences in the shape of the debris are more pronounced, as  
the importance of relativistic effects in a disruption is directly connected 
to the ratio of the periapsis distance $\rp$ to the gravitational radius $\rg$. 
We observe that with all other parameters being the same, the simulations 
with positive BH spin (i.e., of stars on prograde orbits) lie in between the 
Schwarzschild and the Newtonian simulations, with the TDEs with larger 
$\abh$ being closer to Newtonian. Simulations with negative spin (i.e., of stars on retrograde orbits) are more extreme than the Schwarzschild simulations. 
Again, we cannot report a significant difference in the $\dot{M}$ rates for the $\beta=6$ case: the peak $\dot{M}$ values do not vary by more than $\sim 10\%$, and the time at which this maximum is reached does not vary by more than $\sim$ 2 weeks between all simulations. We do note, however, that the $\dot{M}$ curves for the simulations with negative BH spin (dotted lines), and in particular for $\abh=-0.99$, exhibit a rather sharp rise resulting from the wider spread in orbital energies (cf.~Fig.~\ref{fig:espread} for $\beta=6$), which is probably correlated with the ``puffed up'' geometry of the debris stream (see the right panel of Fig.~\ref{fig:spinMdot2}).

At later times (a few months after the periapsis passage), the $\dot{M}$ curves from all of our simulations settle into a $t^{-5/3}$ power law decay curve, which is probably related to our simple initial conditions and stellar structure, and the $\gamma=5/3$ polytropic equation of state. A certain deviation away from the $t^{-5/3}$ decay is expected for stars with more realistic structure, rotation, and more complex EOS. Since the BH spin appears to have very little impact on the shape of the $\dot{M}$ curves, we conclude that it would be extremely difficult to infer it from the fallback rates. 

Our results differ somewhat from \citet{kesden12}, which to our knowledge is the only paper that methodically analyses the influence of BH spin on the return rates, and there are a number of possible sources for this discrepancy. The most prominent is the fact that \citet{kesden12} does not include a self-consistent treatment of the fluid's hydrodynamics or self-gravity. He used a semi-analytical approach that considers the whole stellar material as becoming instantaneously unbound at a predetermined point in its trajectory, namely $\rp$ (though in Sec.~V he mentions that if one were to consider $\rt$ instead, the expected relativistic corrections would be much smaller). However, our simulations do reproduce the qualitative effect of the BH spin, i.e.~the fact that negative spin parameters (in his simulations represented as orbits with inclination $\iota=\pi$) result in higher accretion rates with earlier rise times as compared with $\abh=0$ (see his Figs.~11 and 13, for instance).

\subsection{Impact of spin on the spread in energies after disruption}

The fallback rate $\dot{M}$, the return time for the most bound debris $t_{\rm fall}$, the peak fallback rate $\dot{M}_{\rm peak}$, and the time $t_{\rm Edd}$ at which $\dot{M}$ becomes sub-Eddington all depend on the spread in specific orbital energies $\Delta\en$ of the debris. This spread originates almost entirely from the spread in potential energies across the star at the moment when the star's fluid elements begin moving on geodesic trajectories. This moment has long been considered to be the first periapsis passage of the star around the BH, although more recent studies (e.g.~\mbox{\citealp{guillochon13}}; \mbox{\citealp{stone13}}) have shown that the stellar fluid already starts moving on geodesic trajectories as it enters the tidal radius.

A simple, first-order Taylor expansion of the potential energy at either $\rp$ or $\rt$ yields a very accurate estimate for the energy spread $\Delta\en$, 
\begin{equation}
\Delta\en=k \beta^n \frac{\G\mbh\rst}{\rt^2},
\label{eqspread}
\end{equation} 
where $k$ is a constant of order unity that depends on the stellar structure and rotation. If $\Delta\en$ is computed by taking the potential gradient at the tidal radius, $n=0$ and so $\Delta\en$ is independent of $\beta$, while if $\Delta\en$ is computed at periapsis then $n=2$ (which is the traditional picture). From previous work \mbox{\citep{guillochon13}}, $n$ is expected to have piecewise values or even a more complicated dependence on $\beta$, since disruption does not occur at an instant in time, but is rather gradual, with the stellar material at the surface being stripped away first, and the stellar core being disrupted last; however, as the two estimations are in a sense limiting cases, $n$ is expected to take values between $0$ and $2$.

The relevance of the $n(\beta)$ dependence resides in the observational implications of $n$. We point the reader to Section 8 of \mbox{\citet{stone13}}, where it is shown that $\dot{M}$, $t_{\rm fall}$, $\dot{M}_{\rm peak}$ and $t_{\rm Edd}$ all depend (either linearly or not) on $\beta^n$. Thus, if $n=0$ then none of these quantities will depend on $\beta$, and the signature of a TDE will be completely determined by the combination of $\mbh$, $\mst$ and $\rst$. If, however, $n$ approaches 2, one would expect major differences in the observational signatures of disruptions with different impact parameters, and the number of possible fallback curves would be greatly increased.  Further implications of $n$ for the optical transient searches are discussed by \mbox{\citet{stone13}}.

Given the astrophysical relevance of $\Delta\en$ (and of $n$ itself) for future observations of TDEs, we have used our new method to determine the dependence of $n$ not only on $\beta$, but also on the BH spin. 

For this purpose, we ran a number of simulations ($\mbh=10^6~\msun$, $\mst=\msun$, $\rst=\rsun$) with impact parameters between $\beta=0.5$ and $\beta=11$, and with Newtonian, Schwarzschild and Kerr BHs. For the latter, we considered rotating and counter-rotating BHs with spin parameters $\abh=\pm 0.5$ and $\abh=\pm 0.99$. The simulations were run until 2.5 days (or about 65 dynamical time scales of the initial star) after the first periapsis passage (for $\beta\lesssim 8$), or until the beginning of the second periapsis passage (for the relativistic simulations with $\beta\gtrsim 8$). At the end of the simulations, we computed the mechanical energy $\en$ of each SPH particle, sorted the particles in terms of $\en$, and defined the $\Delta\en$ interval by excluding the 1 per cent of the particles with the lowest energy, and the 1 per cent with the highest, i.e.~$\Delta\en$ is the interval centred on the median energy that contains 98 per cent of all the SPH particles.


\begin{figure}
\centering
\includegraphics[width=\linewidth]{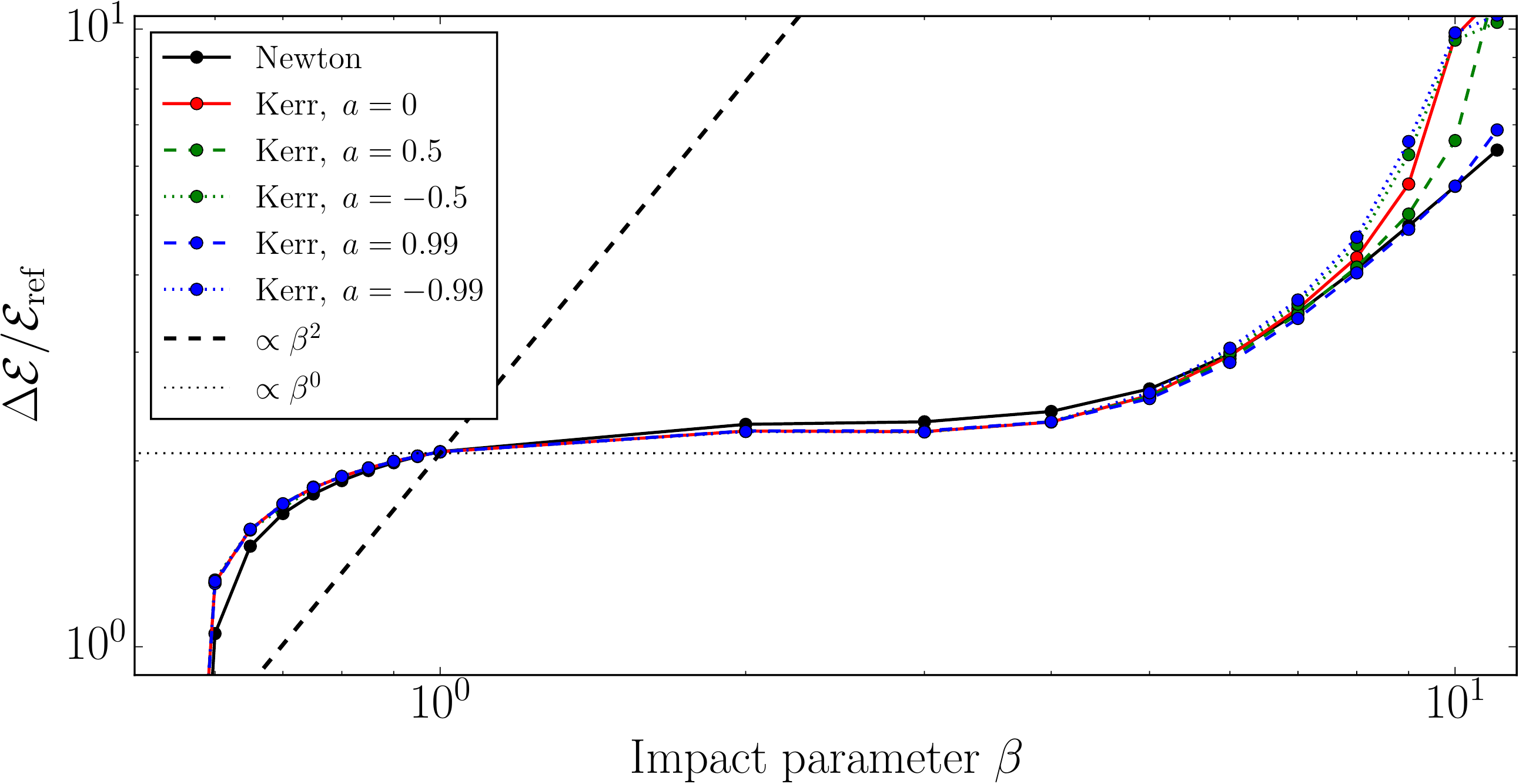}
\caption{Spread in orbital energies $\Delta\en$ after a disruption as a function of the impact parameter $\beta$, in Newtonian simulations (solid black line), and in Kerr simulations with various BH spins: $\abh=0$ (solid red line), and $\pm 0.5$ (dashed/dotted green line) and $\pm 0.99$ (dashed/dotted blue line). The typical $\propto\beta^2$ and $\propto\beta^0$ scalings normally used in analytical studies (see \eqp{eqspread}) are shown with dashed and dotted black lines, respectively. We observe that: a) the BH spin has a negligible effect on $\Delta\en$ below $\beta \simeq 8$, and b) the best fit to the data is given by a piecewise polynomial, with $\Delta\en$ exhibiting a rough empirical scaling $\propto \beta^0$ between $\beta\sim 1$ and $\beta\sim 4$, and $\propto \beta^2$ for larger values of $\beta$.}
\label{fig:espread}
\end{figure}

In Fig.~\ref{fig:espread} we plot the data points for $\Delta\en(\beta)$ as obtained in all six sets of simulations. The latter are differentiated in the plot by line colour and style as described in the plot legend (Newtonian: solid black; Schwarzschild: solid red; prograde/retrograde Kerr: dashed/dotted green and blue, for $\abh=0.5$ and 0.99, respectively).
The spread in energies is normalized by a factor $\en_{\rm ref}$ equal to $ G\mbh\rst/\rt^2$ (see  Eq.~\ref{eqspread}), which means that the constant $k$ can easily be read off the plot, as it is equal to $\Delta\en/\en_{\rm ref}$ at $\beta=1$; thus, for our $\gamma=5/3$ non-rotating polytrope, we find numerically that $k\approx 2.1$.
The figure also contains the analytical fits for $n=0$ and $n=2$ (represented  as dotted and dashed black lines, respectively, and normalized to the same value of $k$ as obtained in the simulations).

In agreement with \citet{stone13}, we find that for such a small $\mbh/\mst$ ratio ($10^6$) relativistic effects are not strong enough to produce any significant deviation from the Newtonian simulations, except for the deepest encounters with impact parameters $\beta\gtrsim 8$. Also, in agreement with both \mbox{\citet{stone13}} and \mbox{\citet{guillochon13}}, we find that for $\beta<1$, in the regime of partial disruptions, $n$ drops sharply, with the cutoff around $\beta\simeq 0.6$, below which the entire star survives undisrupted. For mild encounters ($1 \lesssim \beta \lesssim 4$), $n=0$ is an excellent approximation. For deeper encounters, $n$ increases towards 2 (although in the Newtonian simulations it never reaches $n=2$). This increase in energy spread is accounted for by the extra energy released at the strong shock that forms across the star as it gets more violently compressed for larger values of $\beta$. The fact that this increase is more pronounced for the relativistic encounters is a reflection of the corresponding tidal compression being more severe than the Newtonian one (cf.~discussion in Sec.~\ref{sec:rel_effects}). 

\subsection{Impact of spin on debris geometry}

Finally, we simulated the tidal disruption of a representative white dwarf (WD; $\mst=0.6~\msun$, modelled as a $\gamma=5/3$ polytrope) by a BH with mass $\mbh=10^6~\msun$ and spin parameter $\abh=0.98$. For such a rapidly-rotating BH, the WD can approach as close as $1.2\,\mbh$ without plunging into the BH. 
We used an impact parameter: $\beta=0.5$, corresponding to a ratio $\rp/\rg\simeq 1.5$, and set the star on an inclined orbit (relative to the BH spin), with both the initial and the minimum latitudes (see Appendix~\ref{A2}) equal to $\theta=0.35\,\pi$; the simulation started at a distance of 10~$\rt$ from the BH and was performed in Boyer--Lindquist coordinates.

It is important to stress that such a disruption is only possible for a rotating BH (and therefore it can only be simulated with a code that properly accounts for the fluid motion in Kerr spacetime). In both Newtonian and Schwarzschild simulations, where the accretion radius is normally placed at or outside the Schwarzschild radius ($2\,\mbh$), this combination of orbital parameters would result in the star being promptly swallowed by the BH, without any disruption. 

\begin{figure*}
\centering
\includegraphics[width=\linewidth]{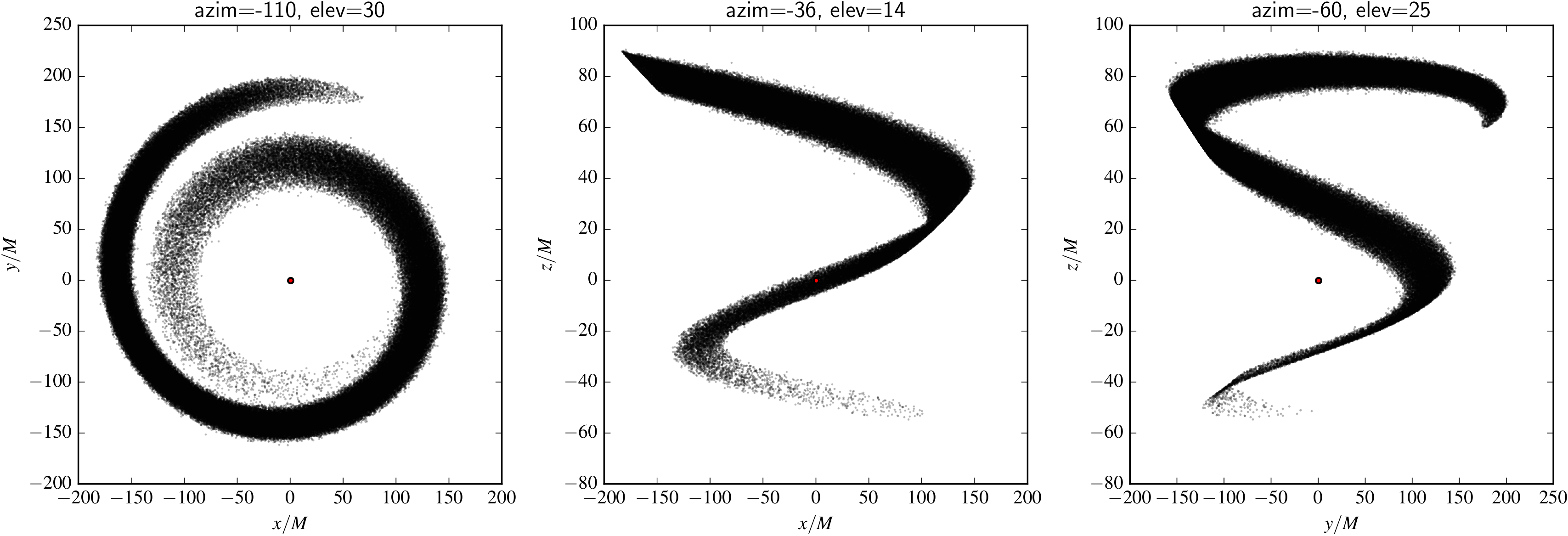}
\includegraphics[width=\linewidth]{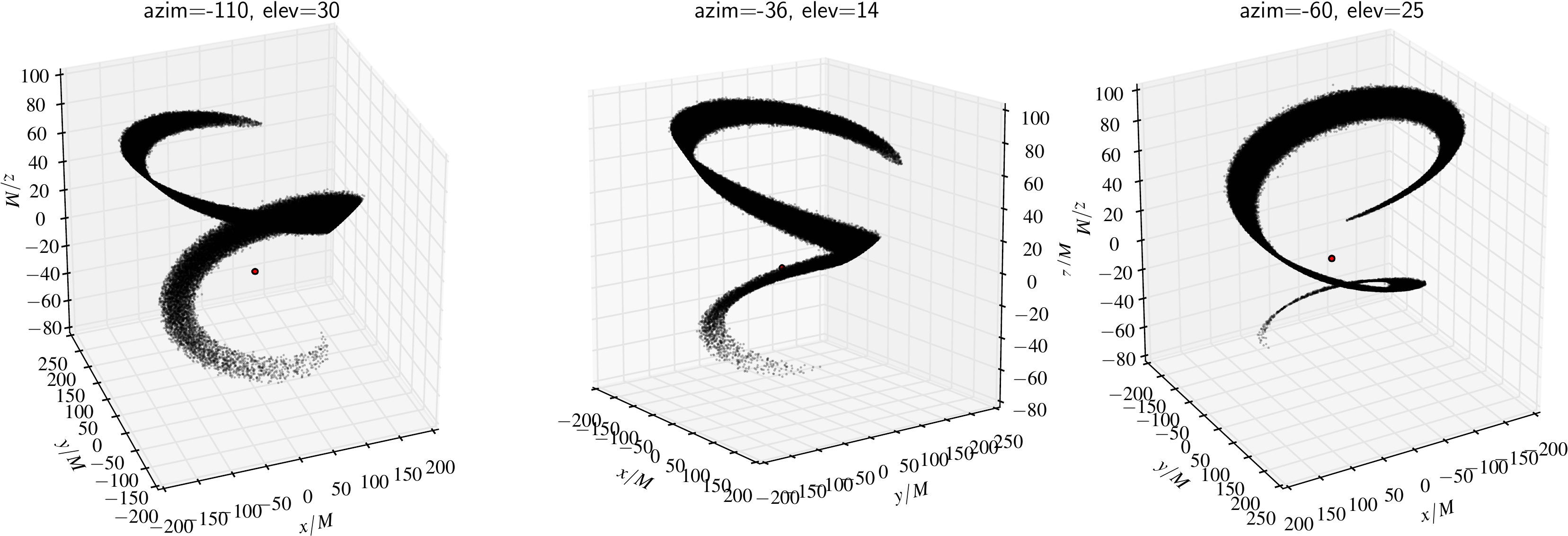}
\caption{Spatial distribution of the debris stream for a close relativistic encounter between a $0.6\,\msun$ white dwarf and a $10^6\,\msun$ BH, with $\rp/\rg\simeq 1.5$. The combined periapsis and orbital plane precession, different for each SPH particle, leads to the debris stream taking a helicoidal shape, while the negligible contribution of self-gravity (even across the stream) leads to the stream thickness increasing with time).}
\label{fig:staircase}
\end{figure*}

The geometrical distribution of the tidal debris is shown in Fig.~\ref{fig:staircase}.
The upper panels present a projection of the SPH particles onto the $x-y$, $x-z$, and $y-z$ planes, while the lower panels present a three-dimensional view of the SPH particles from different perspectives (azimuth and elevation angles of the ``observer'').

First, we find that the WD is completely disrupted in spite of the low impact parameter $\beta$, which can be explained by the strong relativistic effects (increased tidal stresses and extra time spent near periapsis, as discussed in \citealp{GT15}). 
The first of the six plots shows a geometry that is fairly common for a deep relativistic encounter (i.e., with small $\rp/\rg$ ratio), with the different fluid elements experiencing different amounts of periapsis precession, resulting in the debris stream being stretched into a long spiral. In this case, however, the particles also experience individual degrees of orbital plane precession, which results in a helicoidal shape of the debris stream.

We also note that the thickness of the stream increases significantly at later times. Due to the different periapsis shifts experienced by each individual fluid element, not only does the stream take a spiral shape, but -- as long as self-gravity forces are small enough across the stream -- it also increases in thickness. Since the prediction that orbital plane precession will impede the self-intersection of the stream for many orbits is predicated on its being thin, this observed increase in thickness (due to geodesic motion) may have a crucial influence on whether the stream self-intersects early or not, and hence on the circularization time scale. This is an effect that -- to our knowledge -- has not been previously discussed in the literature. It is clear that in order to properly explore this, a thorough exploration of the parameter space is needed. We thus leave this for future studies. We would like to remark, however, that this effect is only to be expected in deep relativistic encounters, i.e.~those for which $\rp < 10\,\rg$.

\section{Summary}
\label{sec:summary}

We have presented here a novel method for modelling relativistic effects in the dynamics of a self-gravitating fluid in the presence of a dominant massive BH. This approximate approach combines an exact relativistic description of the fluid dynamics coupled with a Newtonian treatment of the fluid's self-gravity.  

We have given explicit expressions for the evolution equations relevant for motion around a Kerr BH using both Boyer--Lindquist and Kerr--Schild coordinate systems. We have implemented these equations within a Newtonian SPH code and shown that it performs with  virtually no additional computational cost. 
We have demonstrated the use of this new tool by applying it to the study of stellar tidal disruption events (TDEs) by supermassive BHs. Our approach allows exploration of the star's fate from far away from the BH (in the Newtonian regime) down to the strong relativistic regime near the spinning BH without having to change the simulation methodology. In addition, this approach should also be useful for other astrophysical settings where the mass of a central BH is the dominant factor determining the overall spacetime curvature. 

The new methodology captures to an excellent precision pure geodesic motion, even for off-equatorial orbits and for very deep encounters. Moreover, using this approach we recovered previous results of relativistic simulations of TDEs. As an additional validation test, we have compared the output of several relativistic TDE simulations starting from the same initial conditions but evolved using two different coordinate systems. The resulting centre of mass trajectories and constants of motion deviate from each other by less than a few per cent, but are completely different from the corresponding Newtonian values.

We have applied the new approach to exploring the effect of the BH spin parameter on the fallback rate of the returning debris after a TDE and found that it does not have any significant effect on this. However, the combined effects of periapsis and orbital plane precessions in deep encounters imprint heavily on the morphology of the debris stream. This new tool will be applied in future studies of relativistic effects in tidal disruption encounters.

\section*{Acknowledgements}

We thank James Guillochon, Aleksander S\k{a}dowski, and Illa R.~Losada for useful discussions and comments on the manuscript. We are also grateful to the anonymous referee for the many useful suggestions that have improved our paper after submission. ET acknowledges support from CONACyT grants 290941 and 291113. The work of SR has been supported by the Swedish Research Council (VR) under grant 621-2012-4870. The simulations have been carried out on the facilities of the North-German Supercomputing Alliance (HLRN).

\bibliographystyle{mn2e}

\bibliography{references}\balance

\include{appendix}

\end{document}

%% file: appendix.tex
\onecolumn
\appendix

\setcounter{section}{0}
\setcounter{equation}{0}

\renewcommand{\theequation}{\thesection.\arabic{equation}}

\section{Initial conditions for trajectories in Kerr spacetime}
\label{A}

\subsection{Timelike geodesics and constants of motion}\label{A1}

The line element of the Kerr metric in Boyer--Lindquist (BL) coordinates is 
\begin{equation}
\begin{split}
\ud s^2 & =  -\left( 1 - \frac{ 2Mr }{ \rho^2 } \right)  \ud t^2 
- \frac{ 4\, aM r \sin^2 \theta }{ \rho^2 }\, \ud t\, \ud \phi  \\
& \phantom{=\ } +\frac{ \rho^2 }{ \Delta }\, \ud r^2
+ \rho^2 \ud \theta^2 + \frac{ \Sigma \sin^2 \theta }{ \rho^2 }\, \ud \phi^2.
\end{split}
\label{eA.1}
\end{equation}
with $\rho$, $\Delta$ and $\Sigma$ defined as
\begin{align}
&\rho^2 = r^2 + a^2\cos^2\theta,\nonumber\\ 
&\Delta = r^2 - 2Mr + a^2, \label{eA.2} \\
&\Sigma = \left(r^2 + a^2\right)^2 - a^2 \Delta\sin^2\theta. \nonumber
\end{align}
\noindent Consider a test particle with four-velocity $U^\mu=\ud x^\mu/\ud \tau$ following a time-like geodesic where $\tau$ is the proper time. In addition to the trivially conserved modulus of the four-velocity (i.e.~$U^\mu\,U_\mu=-1$), the symmetries of the Kerr metric lead to the existence of three additional conserved quantities: the specific energy $\mathcal{E}$, the projection of the specific angular momentum along the BH's spin axis $\ell_z$, and the Carter constant $\mathcal{Q}$. Using BL coordinates, these quantities are given by
\begin{gather}
\mathcal{E} = \Gamma
\left(1-\frac{2Mr}{\rho^2} + \frac{2\,aMr\sin^2\theta}{\rho^2}\dot{\phi}\right),\label{eq:A3}\\
\ell_z = \Gamma
\left(\Sigma\dot{\phi} - 2\,aMr\right)\frac{\sin^2\theta}{\rho^2},\label{eq:A4}\\
\mathcal{Q} = \rho^4\Gamma^2\dot\theta^2 + \ell_z^2\cot^2\theta - \varepsilon\,a^2\cos^2\theta,
\end{gather}
where $\varepsilon\equiv\mathcal{E}^2-1$ and $\Gamma$ is the generalized Lorentz factor given by
\begin{equation}
\Gamma =\left(1-\frac{2Mr}{\rho^2}+\frac{4\,aMr\,\sin^2\theta}{\rho^2}\dot{\phi}  -\frac{\rho^2}{\Delta}\dot{r}^2-\rho^2\dot{\theta}^2-\frac{\Sigma\sin^2\theta}{\rho^2}\dot{\phi}^2 \right)^{-\nicefrac{1}{2}}.
\end{equation}
In all of these expressions, an overdot denotes a derivative with respect to the coordinate time $t$. For convenience, we will sometimes also make use of the following combination of the conserved quantities
\begin{equation}
\ell^2 = \mathcal{Q} +(\ell_z-a\,\mathcal{E})^2
\label{ll}
\end{equation}
which is clearly a conserved quantity as well. This quantity is connected to the square of the total magnitude of the angular momentum although being coupled, in a non-trivial way, with the energy of the test particle and the spin of the BH \citep[see][for a discussion about the physical interpretation of $\ell^2$]{defelice99}.

In the case of Kerr spacetime, the existence of these first integrals of motion allows us to reduce the geodesic equations (\eqp{e3.2}) to the following set of partially decoupled,
first-order ordinary differential equations:
\begin{align}
\rho^2 \frac{\ud r }{\ud\tau} & =  \pm \sqrt{\mathcal{R}(r)},
\label{em2}\\
\rho^2 \frac{\ud \theta }{\ud\tau} & =  \pm \sqrt{\Theta(\theta)},
\label{em3}\\
\rho^2 \frac{\ud \phi }{\ud\tau} & =
\frac{\mathcal{A}(\theta)}{\sin^2\theta} +
\frac{a}{\Delta}\,\mathcal{B}(r),
\label{em4}\\
\rho^2 \frac{\ud t }{\ud\tau} & = a\,\mathcal{A}(\theta) +
\frac{r^2+a^2}{\Delta}\,\mathcal{B}(r),
\label{em5}
\end{align}
with
\begin{align}
 \mathcal{R}(r) = &\ \mathcal{B}^2(r)-(r^2+\ell^2)\Delta,
\label{em6}\\
\Theta(\theta) = &\ \mathcal{Q}
+\varepsilon\,a^2\cos^2\theta-\ell_z^2\cot^2\theta,
\label{em7}\\
\mathcal{A}(\theta) = &\ \ell_z-a\,\mathcal{E}\,\sin^2\theta,
\label{em8}\\
\mathcal{B}(r) = &\ \left(r^2+a^2\right)\mathcal{E} - a\,\ell_z.
\label{em9}
\end{align}
The signs in \eqs{em2} and \eqref{em3} are independent of each other
and change whenever the test particle reaches a radial or polar turning point,
respectively, in its trajectory.

\subsection{Initial conditions}\label{A2}

We want to determine the initial velocities for a test particle (or for the centre of mass of an incoming star) given the initial position ($r_0,\theta_0,\phi_0$), the impact parameter $\beta$, and the orbital eccentricity $e$. Note that specifying $\beta$ and $e$ is equivalent to fixing the pericentre $ r_p$ and apocentre $ r_a$ distances of the orbit, since, for a fixed tidal radius $r_\mathrm{t}$, we have
\begin{equation}
 r_p = \frac{r_\mathrm{t}}{\beta}, \qquad  r_a =  r_p\left(\frac{1+e}{1-e}\right).
\end{equation}
For off-equatorial trajectories we also need to provide the angular span of the polar motion, which can be done by either specifying the Carter constant $\mathcal{Q}$ or, perhaps more intuitively, the minimum and maximum latitudes $\theta_a$, $\theta_{a'}$ such that the polar coordinate $\theta$ is restricted to the interval $\theta\in[\theta_a,\,\theta_{a'}]$. Note that these two latitudes are simply related by $\theta_{a'} = \pi-\theta_a$, so it is sufficient to specify one of them, say $\theta_a$.

Our aim now is to find the set of initial velocities $\dot{r}_0$, $\dot{\theta}_0$, $\dot{\phi}_0$ for a given set of initial positions $r_0$, $\theta_0$, $\phi_0$ and turning points $ r_p$,  $ r_a$, $\theta_a$. The corresponding Cartesian positions and velocities can be in turn computed using \eqref{eA.8}-\eqref{eA.10} below. From \eqs{em2}-\eqref{em5} we see that our problem reduces to finding the set of first integrals of motion $\mathcal{E}$, $\ell_z$ and $\mathcal{Q}$ from the information that we already have. From the polar equation \eqref{em3} we have
\begin{equation}
\mathcal{Q} = \ell_z^2\cot^2\theta_a-\varepsilon\,a^2\cos^2\theta_a,
\label{QQ}
\end{equation}
while, from the radial function in \eqref{em6} we get
\begin{equation}
\mathcal{R}(r) = \varepsilon\,r^4+2Mr^3+(\varepsilon\,a^2-\ell_z^2-\mathcal{Q})r^2+2\,\rg\,\ell^2r -a^2\mathcal{Q}.
\label{poly1}
\end{equation}
We can rewrite this expression now in terms of the turning points
\begin{equation}
\mathcal{R}(r) = \varepsilon(r- r_a)(r- r_p)(r-r_c)(r-r_d),
\label{poly2}
\end{equation}
where $r_c$ and $r_d$ are two, as yet unspecified, extra roots of the the polynomial $\mathcal{R}(r)$. If we compare equal powers of $r$ in \eqs{poly1} and \eqref{poly2}, we can solve for the conserved quantities as a function of the turning points as
\begin{gather}
\varepsilon = -\frac{2 M}{ r_a+ r_p+r_c+r_d}, \label{com1}\\
\mathcal{Q} = \frac{2 M  r_a\, r_p\,r_c\,r_d}{a^2( r_a+ r_p+r_c+r_d)},\\
\ell^2 = \frac{ r_a\, r_p(r_c+r_d)+r_c\,r_d( r_a+ r_p)}{ r_a+ r_p+r_c+r_d},\\
\ell_z^2 = 2 M\frac{a^2\left[ r_a\, r_p+r_c\,r_d+( r_a+ r_p)(r_c+r_d)-a^2\right]- r_a\, r_p\,r_c\,r_d}{a^2( r_a+ r_p+r_c+r_d)}. \label{com4}
\end{gather}
We can now combine these equations together with \eqs{ll} and \eqref{QQ} and obtain the following system of two equations in the two unknown roots $r_c$ and $r_d$
\begin{equation}
a^2\cos^2\theta_a\left[ r_a\, r_p+r_c\,r_d+( r_a+ r_p)(r_c+r_d) - a^2\cos^2\theta_a\right] -  r_a\, r_p\,r_c\,r_d = 0,
\label{hle1}
\end{equation}
\begin{equation}
\begin{split}
&\left\{a^2
   ( r_a+ r_p+r_c+r_d-4M)+2M\left[ r_a\, r_p+r_c\,r_d+( r_a+ r_p)(r_c+r_d)\right]- 
    r_a\, r_p(r_c+r_d)-r_c\,r_d( r_a+ r_p)\right\}^2 +  \\
 &\ 8M \left\{ r_a^4+ r_a\, r_p\,r_c\,r_d -
  a^2 \left[ r_a\, r_p+r_c\,r_d+( r_a+ r_p)(r_c+r_d)\right]\right\}( r_a+ r_p+r_c+r_d-2M)=0.
\end{split}   
\label{hle2}
\end{equation}
From \eq{hle1} we can solve for either $r_c$ or $r_d$, and then substitute the result into \eq{hle2}. Doing this gives a fourth-order polynomial in that unknown root. Once this root has been found, we obtain the other one by substituting back into \eq{hle1}. 

For parabolic motion we have that $\varepsilon=0$ and $r_a=\infty$. The rest of the constants of motion are given by
\begin{gather}
\mathcal{Q} = 2 M \frac{ r_p\,r_c\,r_d}{a^2},\\
\ell^2 = \left[ r_p(r_c+r_d)+r_c\,r_d\right],\\
\ell_z^2 = 2 M\left[ r_p+r_c+r_d-\frac{ r_p\,r_c\,r_d}{a^2}\right].
\end{gather}
 In this case the equations to solve are
\begin{equation}
r_c\,r_d\, r_p-a^2\,\cos^2\theta_a(r_c+r_d+ r_p)=0,
\label{lhle1}
\end{equation}
\begin{equation}
\left[a^2+2M( r_p+r_c+r_d)- r_p(r_c+r_d)-r_c\,r_d\right]^2 
+ 8M\left[ r_p\,r_c\,r_d -a^2( r_p+r_c+r_d)\right]=0.
\label{lhle2}
\end{equation}
Analogously to the previous case, from \eq{lhle1} we can solve for either $r_c$ or $r_d$, and then substitute the result into \eq{lhle2}. Doing this results now in a third-order polynomial in that unknown root. Once this root has been found, we obtain the other one by substituting back into \eq{lhle1}. 

With all of the turning points at hand, we can now compute the corresponding constants of motion $\varepsilon$, $\mathcal{Q}$, $\ell$, and $\ell_z$ using \eqs{com1}-\eqref{com4}. Next, the initial velocities are computed using \eqs{em2}-\eqref{em5}, which, in turn, can be transformed into Cartesian-like coordinates and velocities via \eq{eA.3} and \eqs{eA.8}-\eqref{eA.10}.

\section{Cartesian form of the Boyer--Lindquist coordinates}
\label{AA}

The Cartesian-like coordinates $(x,\,y,\,z)$ associated with the spatial BL coordinates $(r, \theta, \phi)$ are defined as
\begin{align}
&x  = \sqrt{r^2+a^2}\,\sin\theta\,\cos\phi,\nonumber\\ 
&y  = \sqrt{r^2+a^2}\,\sin\theta\,\sin\phi,\label{eA.3} \\
&z  = r\,\cos\theta, \nonumber
\end{align}
while the time coordinate $t$ is taken to be the same in both systems. By inverting \eq{eA.2} we get the following expressions for the inverse transformation
\begin{align}
&r = \sqrt{ \frac{1}{2}\left(x^2 + y^2 + z^2 - a^2\right) +
\frac{1}{2}\sqrt{\left(x^2 + y^2 + z^2 - a^2\right)^2 + 4\, a^2 z^2}},
\nonumber\\
&\theta = \cos^{-1}\left(\frac{z}{r}\right),\label{eA.4}\\
&\psi = \tan^{-1}\left(\frac{y}{x}\right).\nonumber
\end{align}

\noindent In terms of Cartesian-like coordinates, the differential line element is
\begin{equation}
\begin{split}
\ud s^2 = & - (\cc\,\ud t)^2 + \ud x^2 + \ud y^2 + \ud z^2  \\
&+\frac{2M r}{\rho^2}\left\{\left[dt-a\left(\frac{x\,\ud y-y\,\ud x}{r^2+a^2}\right)\right]^2
+\frac{\left[ r^2\left(x\,\ud x+y\,\ud y\right)+(r^2+a^2)z\,\ud z \right]^2}{r^2\Delta(r^2+a^2)}
\right\},
\end{split}\label{eA.5}
\end{equation}
where now $r$ and $\rho$ should be considered as implicit functions of $(x,\,y,\,z)$ satisfying
\begin{gather}
r^4-r^2(x^2+y^2+z^2-a^2)-a^2z^2 = 0, \label{eA.6}\\
\rho^2 = r^2+\frac{a^2z^2}{r^2} . \label{eA.7}
\end{gather}

\subsection{Velocity transformations}

By differentiating \eq{eA.3} with respect to the coordinate time $t$ we obtain the following expressions for the velocity transformations
\begin{gather}
\dot{x} = \frac{r}{\sqrt{r^2+a^2}}\,\sin\theta\,\cos\phi\,\dot{r} + \sqrt{r^2+a^2}\left(\cos\theta\,\cos\phi\,\dot{\theta} -\sin\theta\,\sin\phi\,\dot{\phi}\right), \label{eA.8}\\
\dot{y} = \frac{r}{\sqrt{r^2+a^2}}\,\sin\theta\,\sin\phi\,\dot{r} + \sqrt{r^2+a^2}\left(\cos\theta\,\sin\phi\,\dot{\theta} + \sin\theta\,\cos\phi\,\dot{\phi}\right), \label{eA.9}\\
\dot{z} = \cos\theta\,\dot{r}-r\,\sin\theta\,\dot{\theta},\label{eA.10}
\end{gather}
where a dot denotes differentiation with respect to $t$. We can invert these expressions as
\begin{gather}
\rho^2\dot{r}  = r\left(x\,\dot{x}+y\,\dot{y}+z\,\dot{z}\right)+\frac{a^2z\,\dot{z}}{r}, 
\label{eA.11} \\
\rho^2\dot{\theta}  = \frac{z\left(x\,\dot{x}+y\,\dot{y}+z\,\dot{z}\right)-r^2\dot{z}}{\sqrt{r^2-z^2}},
\label{eA.12}\\
\dot{\phi}  = \frac{x\,\dot{y}-y\,\dot{x}}{x^2+y^2}.
\label{eA.13}
\end{gather}

\subsection{Constants of motion}

In this section we collect the necessary expressions for calculating the constants of motion from the Cartesian-like coordinates.
\begin{equation}
\mathcal{E} = \Gamma \left\{ 1 - \frac{ 2Mr}{\rho^2}\left[1-a\left(\frac{x\,\dot{y}-y\,\dot{x}}{r^2+a^2}\right)\right] \right\},
\end{equation}
\begin{equation}
\ell_z = \Gamma \left\{x \,\dot{y} -  \dot{x}\, y - \frac{2Ma\,r}{\rho^2} \left(\frac{x^2+y^2}{r^2+a^2}\right) \left[1-a\left(\frac{x\,\dot{y}-y\,\dot{x}}{r^2+a^2}\right)\right] \right\},
\end{equation}
\begin{equation}
\ell^2 = \Gamma^2 \left[r^2\left(\dot{x}^2+\dot{y}^2+\dot{z}^2\right) - \left(x\,\dot{x}+y\,\dot{y}+z\,\dot{z}\right)^2
 - 2\,a(x\,\dot{y}-y\,\dot{x}) + a^2\left(\frac{x^2+y^2}{r^2+a^2}+\dot{x}^2+\dot{y}^2\right)\right]+\frac{a^2z^2}{r^2}.
\end{equation}
The Lorentz factor $\Gamma=\ud t/\ud \tau$ is calculated from the expression for the metric in \eq{eA.4} as
\begin{equation}
\Gamma = \left(1-\dot{x}^2-\dot{y}^2-\dot{z}^2-
\frac{2Mr}{\rho^2}\left\{\left[1-a\left(\frac{x\,\dot{y}-y\,\dot{x}}{r^2+a^2}\right)\right]^2
+\frac{\left[ r^2\left(x\,\dot{x}+y\,\dot{y}\right)+(r^2+a^2)z\,\dot{z} \right]^2}{r^2\Delta(r^2+a^2)}
\right\}\right)^{-1/2}.
\end{equation}

\subsection{Acceleration}

The hydrodynamic terms in the evolution equation are given by 
\begin{align}
\Gamma^2\varrho\,\omega\,\ddot{x} & = -\frac{\partial P}{\partial x} - 
\frac{\partial P}{\partial t}\left\{ \dot{x}\left[1+\frac{2Mr(r^2+a^2)}{\Delta\,\rho^2}\right]+\frac{2\,aMr\,y}{\Delta\,\rho^2}\right\}  \\
& \hspace{12pt} + \frac{\partial P}{\partial x} \left(\frac{2Mr}{\rho^2}\right)\left[ \frac{1}{r^2+a^2}\left(\frac{r^2x^2}{r^2+a^2}+\frac{a^2y^2}{\Delta}\right) + \frac{a\,y\,\dot{x}}{\Delta}\right] \nonumber \\
& \hspace{12pt} + \frac{\partial P}{\partial y} \left(\frac{2Mr}{\rho^2}\right)\left[ \frac{x\,y}{r^2+a^2}\left(\frac{r^2}{r^2+a^2}-\frac{a^2}{\Delta}\right) - \frac{a\,x\,\dot{x}}{\Delta} \right] \nonumber \\
& \hspace{12pt} + \frac{\partial P}{\partial z}\left(\frac{2Mr}{\rho^2}\right)\frac{x\,z}{r^2+a^2} , \nonumber \\
\Gamma^2\varrho\,\omega\,\ddot{y} & = -\frac{\partial P}{\partial y} - 
\frac{\partial P}{\partial t}\left\{ \dot{y}\left[1+\frac{2Mr(r^2+a^2)}{\Delta\,\rho^2}\right]-\frac{2\,aMr\,x}{\Delta\,\rho^2}\right\}  \\
& \hspace{12pt} + \frac{\partial P}{\partial x} \left(\frac{2Mr}{\rho^2}\right)\left[ \frac{x\,y}{r^2+a^2}\left(\frac{r^2}{r^2+a^2}-\frac{a^2}{\Delta}\right) + \frac{a\,y\,\dot{y}}{\Delta} \right] \nonumber \\
& \hspace{12pt} + \frac{\partial P}{\partial y} \left(\frac{2Mr}{\rho^2}\right)\left[ \frac{1}{r^2+a^2}\left(\frac{r^2y^2}{r^2+a^2}+\frac{a^2x^2}{\Delta}\right) - \frac{a\,x\,\dot{y}}{\Delta}\right] \nonumber \\
& \hspace{12pt} + \frac{\partial P}{\partial z}\left(\frac{2Mr}{\rho^2}\right)\frac{y\,z}{r^2+a^2} , \nonumber \\
\Gamma^2\varrho\,\omega\,\ddot{z} & = -\frac{\partial P}{\partial z} - 
\frac{\partial P}{\partial t}\left\{ \dot{z}\left[1+\frac{2Mr(r^2+a^2)}{\Delta\,\rho^2}\right]\right\}  \\
& \hspace{12pt} + \frac{\partial P}{\partial x} \left(\frac{2Mr}{\rho^2}\right)\left(\frac{x\,z}{r^2+a^2}+\frac{a\,y\,\dot{z}}{\Delta}\right) \nonumber \\
& \hspace{12pt} + \frac{\partial P}{\partial y} \left(\frac{2Mr}{\rho^2}\right)\left(\frac{y\,z}{r^2+a^2}-\frac{a\,x\,\dot{z}}{\Delta}\right) \nonumber \\
& \hspace{12pt} + \frac{\partial P}{\partial z}\left(\frac{2Mr}{\rho^2}\right)\frac{z^2}{r^2} . \nonumber 
\end{align}

\section{Cartesian form of the Kerr--Schild coordinates}
\label{AB}

The Kerr--Schild (KS) system of coordinates is connected to the BL one through the transformation
\begin{equation}
\begin{split}
\ud T    & = \ud t    +\frac{2Mr}{\Delta} \ud r,\\
\ud \psi & = \ud \phi +\frac{a}{\Delta}   \ud r.
\end{split}
\label{eB.1}
\end{equation}

\noindent Using this transformation, the line element of the Kerr metric in KS coordinates becomes 
\begin{equation}
\begin{split}
\ud s^2  = & -\left( 1 - \frac{ 2Mr }{ \rho^2 } \right)  \ud T^2 + \left( 1 +
\frac{ 2Mr }{ \rho^2 } \right)  \ud r^2
  +\frac{4Mr}{\rho^2}\,\ud r\,\ud T- \frac{4Ma\,r}{\rho^2}\,\sin^2 \theta \, \ud
\psi\, \ud T \\ 
  & -2\,a\left( 1 + \frac{ 2Mr }{ \rho^2 } \right)\sin^2
\theta\,\ud r\,\ud
\psi+ \rho^2\,\ud\theta^2+\frac{ \Sigma \sin^2 \theta }{ \rho^2 }\, \ud \psi^2.
\end{split}
\label{eB.2}
\end{equation}
The Cartesian-like coordinates $(x,\,y,\,z)$ associated with the spatial KS coordinates $(r, \theta, \psi)$ are defined as
\begin{align}
&x  = \sin\theta (r\,\cos\psi-a\,\sin\psi),\nonumber\\ 
&y  = \sin\theta (r\,\sin\psi+a\,\cos\psi),\label{eB.3} \\
&z  = r\,\cos\theta, \nonumber
\end{align}
while the time coordinate $T$ is taken to be the same in both systems.

In terms of these coordinates, the differential line element is
\begin{equation}
\ud s^2 =  - \ud T^2 + \ud x^2 + \ud y^2 + \ud z^2  
+\frac{2Mr}{\rho^2}\left[\ud T+ \frac{(r\,x+a\,y)\ud x + (r\,y-a\,x)\ud y}{r^2+a^2}
+\frac{z\,\ud z}{r}\right]^2 .
\label{eB.4}
\end{equation}

\subsection{Velocity transformations}

By differentiating \eq{eB.3} with respect to the coordinate time $T$, we obtain the following expressions for the velocity transformations
\begin{gather}
\dot{x} = \sin\theta\,\cos\psi\,\dot{r} + (r\,\cos\psi-a\,\sin\psi)\cos\theta\,\dot{\theta} -(r\,\sin\psi+a\,\cos\psi)\sin\theta\,\dot{\psi},\\
\dot{y} = \sin\theta\,\sin\psi\,\dot{r} + (r\,\sin\psi+a\,\cos\psi)\cos\theta\,\dot{\theta} +(r\,\cos\psi-a\,\sin\psi)\sin\theta\,\dot{\psi}, \\
\dot{z} = \cos\theta\,\dot{r}-r\,\sin\theta\,\dot{\theta},
\end{gather}
where a dot denotes differentiation with respect to $T$. We can invert these expressions as
\begin{gather}
\rho^2\dot{r}  = r\left(x\,\dot{x}+y\,\dot{y}+z\,\dot{z}\right)+\frac{a^2z\,\dot{z}}{r},  \\
\rho^2\dot{\theta}  = \frac{z\left(x\,\dot{x}+y\,\dot{y}+z\,\dot{z}\right)-r^2\dot{z}}{\sqrt{r^2-z^2}},\\
\dot{\psi}  = \frac{x\,\dot{y}-y\,\dot{x}}{x^2+y^2} + \frac{a\,r}{\rho^2}\left(\frac{x\,\dot{x}+y\,\dot{y}}{r^2+a^2}+\frac{z\,\dot{z}}{r^2}\right).
\end{gather}

\subsection{Constants of motion}

\begin{equation}
\mathcal{E} = \Gamma \left\{ 1 - \frac{ 2\,M\,r}{\rho^2}\left[1+\frac{\dot{x}(r\,x+a\,y)+\dot{y}(r\,y-a\,x)}{r^2+a^2}+\frac{z\,\dot{z}}{r}\right]  \right\},
\end{equation}
\begin{equation}
\ell_z = \Gamma \left\{ x\, \dot{y} - y\,\dot{x} - \frac{2\,M\,a\,r \left(x^2+y^2\right)}{\rho^2
   \left(r^2+a^2\right)} \left[1+\frac{\dot{x}(r\,x+a\,y)+\dot{y}(r\,y-a\,x)}{r^2+a^2}+\frac{z\,\dot{z}}{r} \right] \right\},
\end{equation}
\begin{equation}
\ell^2 = \Gamma^2 \left[r^2\left(\dot{x}^2+\dot{y}^2+\dot{z}^2\right) - \left(x\,\dot{x}+y\,\dot{y}+z\,\dot{z}\right)^2
 - 2\,a(x\,\dot{y}-y\,\dot{x}) + a^2\left(\frac{x^2+y^2}{r^2+a^2}+\dot{x}^2+\dot{y}^2\right)\right]+\frac{a^2z^2}{r^2}.
\end{equation}
where the Lorentz factor $\Gamma$ is calculated from the the expression of the metric in \eq{eB.4} as
\begin{equation}
\Gamma = \left\{1-\dot{x}^2-\dot{y}^2-\dot{z}^2-
\frac{2\,M\,r}{\rho^2}\left[1+\frac{\dot{x}(r\,x+a\,y)+\dot{y}(r\,y-a\,x)}{r^2+a^2}+\frac{z\,\dot{z}}{r}\right]^2\right\}^{-1/2}.
\end{equation}

\subsection{Acceleration}

The hydrodynamic terms in the evolution equation are given by 
\begin{align}
\ddot{x} & = -\frac{1}{\Gamma^2\varrho\,\omega} \left[\frac{\partial P}{\partial x} + \dot{x}\,\frac{\partial P}{\partial t} + \frac{2Mr}{\rho^2}\left(\dot{x} +  \frac{r\,x+a\,y}{r^2+a^2}\right)\mathcal{A}\right], \\
\ddot{y} & = -\frac{1}{\Gamma^2\varrho\,\omega} \left[\frac{\partial P}{\partial y} + \dot{y}\,\frac{\partial P}{\partial t} + \frac{2Mr}{\rho^2}\left(\dot{y} +  \frac{r\,y-a\,x}{r^2+a^2}\right)\mathcal{A}\right], \\
\ddot{z} & = -\frac{1}{\Gamma^2\varrho\,\omega} \left[\frac{\partial P}{\partial z} + \dot{z}\,\frac{\partial P}{\partial t} + \frac{2Mr}{\rho^2}\left(\dot{z} +  \frac{z}{r}\right)\mathcal{A}\right] 
\end{align}
where 
\begin{equation}
\mathcal{A}= \frac{\partial P}{\partial t} - \frac{\partial P}{\partial x}\left(\frac{r\,x+a\,y}{r^2+a^2}\right) - \frac{\partial P}{\partial y}\left(\frac{r\,y-a\,x}{r^2+a^2}\right)-\frac{\partial P}{\partial z} \frac{z}{r} .
\end{equation}

\label{lastpage}